\documentclass[letterpaper,10pt,twoside]{article}

\usepackage{hyperref}
\usepackage{lipsum}
\usepackage[LGRgreek]{mathastext}
\usepackage{scrextend} 
\usepackage{booktabs}
\usepackage{array} 
\usepackage{natbib}
\usepackage{graphicx}
\usepackage{amsmath}
\usepackage{amssymb}

\usepackage[margin=0.6in]{geometry}
\usepackage{fancyhdr}
\newcommand\blfootnote[1]{%
  \begingroup
  \renewcommand\thefootnote{}\footnote{#1}%
  \addtocounter{footnote}{-1}%
  \endgroup
}

\begin{document}




\begin{center}{ 
{\footnotesize Accepted to the \emph{Journal of Astronomical Instrumentation} on 11$^t$$^h$ November 2018 \\ SOFIA Special Edition}

{\large \bf \vspace{1.6cm} SOFIA - HIRMES: Looking forward to the HIgh-Resolution Mid-infrarEd Spectrometer}

\vspace{1cm}

Samuel~N.~Richards$^{1}$$^{\star}$\blfootnote{{\href{mailto:samuel.n.richards@nasa.gov}{$^{\star}$samuel.n.richards@nasa.gov}}}, 
Samuel~H.~Moseley$^{2}$,
Gordon~Stacey$^{3}$,
Matthew~Greenhouse$^{2}$, 
Alexander~Kutyrev$^{2}$,
Richard~Arendt$^{2}$,
Hristo~Atanasoff$^{2}$,
Stuart~Banks$^{2}$,
Regis~P.~Brekosky$^{2}$,
Ari-David~Brown$^{2}$,
Berhanu~Bulcha$^{2}$,
Tony~Cazeau$^{2}$,
Michael~Choi$^{2}$,
Felipe~Colazo$^{2}$,
Chuck~Engler$^{2}$, 
Theodore~Hadjimichael$^{2}$,
James~Hays-Wehle$^{2}$,
Chuck~Henderson$^{4}$,
Wen-Ting~Hsieh$^{2}$,
Jeffrey~Huang$^{5,6}$,
Iver~Jenstrom$^{2}$,
Jim~Kellogg$^{2}$,
Mark~Kimball$^{2}$,
Attila~Kov{\'a}cs$^{8}$,
Steve~Leiter$^{2}$,
Steve~Maher$^{2}$,
Robert~McMurray$^{5}$,
Gary~J.~Melnick$^{9}$,
Eric~Mentzell$^{2}$, 
Vilem~Mikula$^{10}$,
Timothy~M.~Miller$^{2}$,
Peter~Nagler$^{2}$,
Thomas~Nikola$^{4}$,
Joseph~Oxborrow$^{11}$,
Klaus~M.~Pontoppidan$^{12}$,
Naseem~Rangwala$^{13}$, 
Alan~Rhodes$^{5}$,
Aki~Roberge$^{14}$,%
Stefan~Rosner$^{7}$,
Karwan~Rostem$^{2}$,
Nancy~Rustemeyer$^{5}$,
Elmer~Sharp$^{2}$,
Leroy~Sparr$^{2}$,
Dejan~Stevanovic$^{1}$,
Peter~Taraschi$^{2}$,
Pasquale~Temi$^{15}$,
William~D.~Vacca$^{1}$,
Jordi~Vila~Hernandez~de~Lorenzo$^{2}$,
Bill~Wohler$^{5,7}$,
Edward~J.~Wollack$^{2}$,
Shannon~Wilks$^{2}$}

\vspace{0.5cm}

{\footnotesize
$^{1}$
SOFIA Science Center, USRA, NASA Ames Research Center, M/S 232-12, P.O. Box 1, Moffett Field, CA 94035, USA\\
$^{2}$
NASA Goddard Space Flight Center, 8800 Greenbelt Rd, Greenbelt, MD 20771, USA\\
$^{3}$
Department of Astronomy, Cornell University, Ithaca, NY 14853, USA\\ 
$^{4}$
Cornell Center for Astrophysics and Planetary Science, Cornell University, Ithaca, NY, 14853, USA\\ 
$^{5}$
NASA Ames Research Center, M/S 232-12, P.O. Box 1, Moffett Field, CA 94035, USA\\
$^{6}$
LOGYX LLC, Mountain View, CA 94043, USA\\
$^{7}$
SETI Institute, Mountain View, CA 94043, USA\\
$^{8}$
Smithsonian Astrophysical Observatory Submillimeter Array (SMA), MS-78, 60 Garden St, Cambridge, MA 02138, USA\\
$^{9}$
Harvard-Smithsonian Center for Astrophysics, 60 Garden St., Cambridge, MA 02138, USA\\
$^{10}$ 
American University, Institute for Integrated Space Studies \& Technology, Dept. of Physics, Don Myers Building, 4400 Massachusetts Avenue NW, Washington, DC 20016-8079, USA\\ 
$^{11}$
Scientific and Biomedical Microsystems, 806 Cromwell Park Drive, Suite R, Glen Burnie, MD 21061, USA\\
$^{12}$
Space Telescope Science Institute, 3700 San Martin Drive, Baltimore, MD 21218, USA\\
$^{13}$
Space Science and Astrobiology Division, NASA Ames Research Center, Moffett Field, CA 94035, USA\\
$^{14}$
Exoplanets and Stellar Astrophysics Lab, NASA Goddard Space Flight Center, Greenbelt, MD 20771, USA\\
$^{15}$ 
Astrophysics Branch, NASA Ames Research Center, Moffett Field, CA 94035, USA\\
}

\vspace{0.3cm} 

\end{center}

\begin{addmargin}[1cm]{1cm}
{\noindent \small
The HIgh-Resolution Mid-infrarEd Spectrometer (HIRMES) is the 3$^{rd}$ Generation Instrument for the Stratospheric Observatory For Infrared Astronomy (SOFIA), currently in development at the NASA Goddard Space Flight Center (GSFC), and due for commissioning in 2019. By combining direct-detection Transition Edge Sensor (TES) bolometer arrays, grating-dispersive spectroscopy, and a host of Fabry-Perot tunable filters, HIRMES will provide the ability for High Resolution (R$\sim$100,000), Mid-Resolution (R$\sim$10,000), and Low-Resolution (R$\sim$600) slit-spectroscopy, and 2D Spectral Imaging (R$\sim$2000 at selected wavelengths) over the 25 -- 122 $\mu$m mid-far infrared waveband. The driving science application is the evolution of proto-planetary systems via measurements of water-vapor, water-ice, deuterated hydrogen (HD), and neutral oxygen lines. However, HIRMES has been designed to be as flexible as possible to cover a wide range of science cases that fall within its phase-space, all whilst reaching sensitivities and observing powers not yet seen thus far on SOFIA, providing unique observing capabilities which will remain unmatched for decades. }
\end{addmargin} 

\vspace{0.4cm}   

\begin{addmargin}[1cm]{1cm}
\end{addmargin}

\thispagestyle{empty}
\section{Introduction}\label{Section:Introduction}

As all astronomical instrumentation should be driven by the scientific demand, such is the case for the HIgh-Resolution Mid-infrarEd Spectrometer (HIRMES), planned for commissioning on the Stratospheric Observatory For Infrared Astronomy \citep[SOFIA; ][]{2014ApJS..212...24T} in 2019 as a facility-class instrument. HIRMES is currently undergoing its Integration and Testing (I\&T) phase of its development at the NASA Goddard Space Flight Center (GSFC) in partnership with Cornell University, with Samuel~H.~(Harvey)~Moseley as the Principle Investigator. In September 2016, HIRMES was selected to be SOFIA's 3$^{rd}$ Generation Instrument. Since then, it has been following an intensive build schedule, working closely with the NASA/SOFIA Science Instrument Development Team at the NASA Ames Research Center (ARC). 

\vspace{0.5cm}

Following the goal of understanding of how the Earth obtained its water, HIRMES will provide answers to fundamental questions currently being asked in proto-planetary science. Among these questions, HIRMES will tackle: (a) How does the mass of the disk evolve during planetary formation? (b) What is the distribution of oxygen, water ice, and water vapor in different phases of planet formation? (c) What are the kinematics of water vapor and oxygen in proto-planetary disks? In answering these questions, HIRMES will discover where, and in what form, the raw materials for life reside, and how planetary systems like our own evolve. 

HIRMES will quantitatively answers these questions by providing low (R$\sim$600) to very high (R$\sim$100,000) spectral resolving power over the critical spectral range of 25 -- 122 $\mu$m in the mid- to far-infrared waveband. HIRMES combines grating dispersive spectroscopy and Fabry-Perot tunable narrow-band filters with high efficiency background-limited direct detectors. The instrument spectral resolution is designed to match the width of the spectral lines, significantly reducing the background noise, to achieve the maximum possible sensitivity for mid-far infrared spectroscopy with SOFIA. Providing this combination of sensitivity and spectral resolution, HIRMES will open up a unique and useful window on the evolution of planetary systems to the astronomical community. HIRMES' order-of-magnitude sensitivity improvement to SOFIA's current capabilities is crucial for this science program, increasing the number of observable Solar mass proto-planetary systems from a couple to hundreds. Furthermore, the instrument has utility far beyond the aforementioned  investigations, providing tools for a range of Galactic studies, such as stellar outflows and their impact on the interstellar medium, and extragalactic studies, such as the strength and shape of important diagnostic emission lines. 

The instrument's grating mode spectroscopy is a powerful tool for the study of water ice emission in a wide range of objects, an important capability largely absent for more than 20 years since the Kuiper Airborne Observatory (KAO) and Infrared Space Observatory (ISO).

\subsection{Science drivers}\label{Subsection:Science_Drivers}

\begin{figure}
\centering
\centerline{\includegraphics[width=\linewidth]{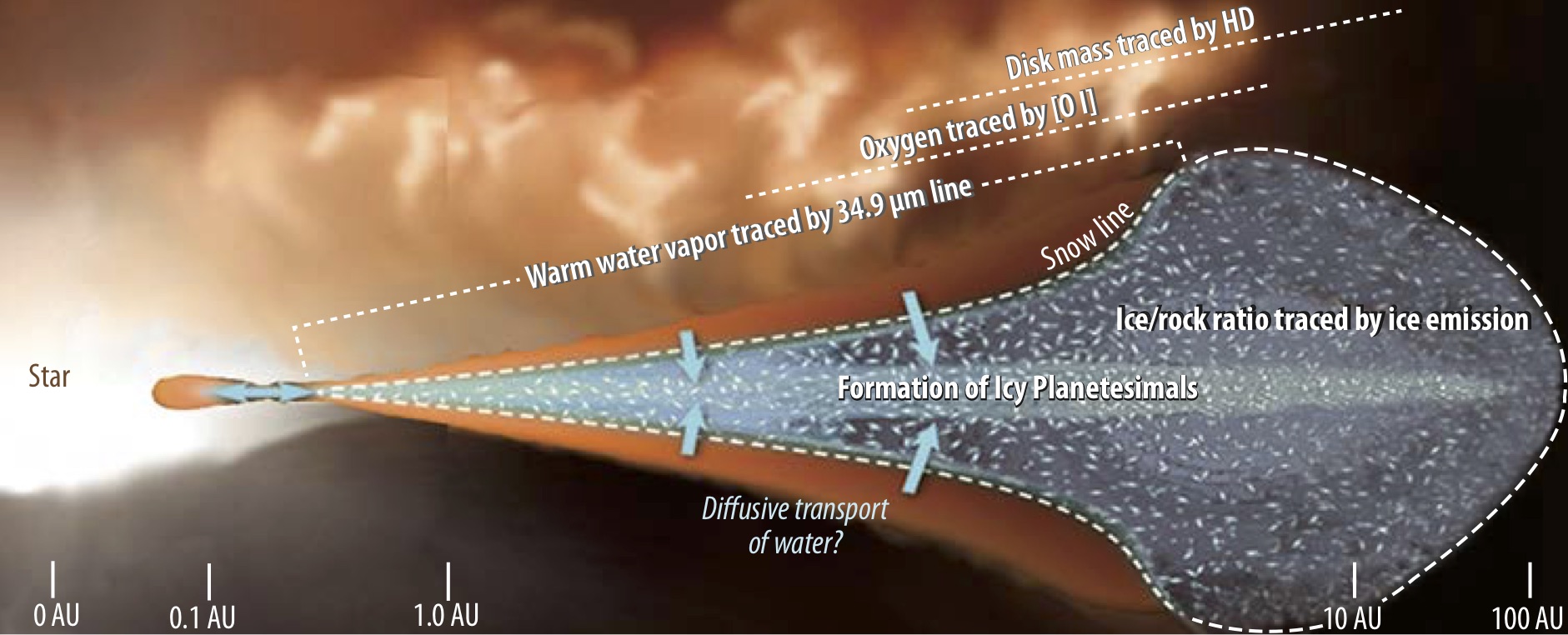}}
\caption{An artistic representation of a proto-planetary disk, sliced edge-on, with a size scale similar to that of our Solar System [Mercury $\simeq$ 0.4 AU, Earth = 1 AU, Saturn $\simeq$ 9.5 AU]. The temperature decreases with distance from the star and with thermal shielding, so any water present transitions through its various states accordingly. Analyzing the three key spectral features shown here in High-Resolution; H$_2$0, [OI] and HD, allows HIRMES to probe different regions of disks and their properties.}
\label{fig:PPD}
\end{figure}

The past decade has seen considerable advancements in our understanding of exoplanets: their diverse properties (e.g. masses, size, orbits) and fundamental trends (e.g. with metallicity or host mass). These observed properties place important boundary conditions on the key processes that govern the formation and evolution of planetary systems. Analogous studies of the proto-planetary disks orbiting young stars provide the vital initial conditions for making planetary systems, as well as the early coevolution and interaction of planets and their birth environments. The observed properties of both disks and exoplanets are needed to inform, test, and refine models of the planet formation process \citep[][and references therein]{2015PASP..127..961A}. Studying the dynamics and chemistry of molecular and atomic gas in the inner regions of the disk (Radius $<$10 AU) provides key information on the reservoir available for the formation of gas giants, and the generation and eventual delivery of such chemicals to terrestrial planets \citep{2013ChRv..113.9016H,2006PNAS..10312249V}. ISO, Spitzer, and Herschel pioneered the studies of proto-planetary disks at mid-far infrared to sub-millimeter wavelengths. SOFIA, in synergy with other ground and space based observatories, will revolutionize this field over the next decade, and thus not only provide the astronomical community with unique data, but also drive the science requirements and design of future missions like the Origins Space Telescope \citep[OST;][]{2018NatAs...2..596B}.

Development of ideas about the role of water in disk surfaces is currently based on limited data, taken at low spectral resolution, or based upon upper limits. The critical transition region from the “wet” inner disk to the “dry” outer disk at a few to 10 AU is beyond our current observational reach. Wavelengths shorter than $\sim$30 $\mu$m (accessible with JWST) trace the innermost disk at $\leq$1 AU, while Herschel traced the cooler outer disk at distances $>$1 AU. However, neither of these observatories spectrally resolve the molecular lines nor do they provide information about the location of the emission. While HIRMES cannot observe the water lines tracing the dilute water vapor beyond the snow line at temperature $<$150 K, observing from the stratosphere on-board SOFIA does allow HIRMES to observe water lines tracing gas at 200 -- 300 K. By spectrally resolving the lines, HIRMES can map the surface abundance of water inside of the transition region. This is achieved by targeting three key spectral lines, specifically: H$_2$O : 34.9 $\mu$m, [OI] : 63.1 $\mu$m, and HD : 112 $\mu$m (see Figure \ref{fig:PPD}). However, there are hundreds of potentially bright water lines accessible to HIRMES over its bandwidth. These three key lines are selected as they probe different regions and properties of the proto-planetary disk, and when combined, they break the degeneracy between mass and abundance that is present in other observational techniques, enabling determination of the timescales over which water is implanted into icy bodies to seed habitable worlds. With this, the proto-planetary disk gas mass, seen as the most fundamental quantity that determines whether planets can form, is now within the grasp of observational astrophysics. Furthermore, HD has been proposed as the best tracer of the total H2 gas mass in proto-planetary disks \citep{2013Natur.493..644B,2016ApJ...831..167M}. HIRMES provides unique access to this mass tracer, which can in turn be used to derive precise molecular abundances in disks, including for observations obtained with ALMA \citep{2017ASSL..445....1B,2018ApJ...865..155C}.

\begin{figure}
\centering
\centerline{\includegraphics[width=10cm]{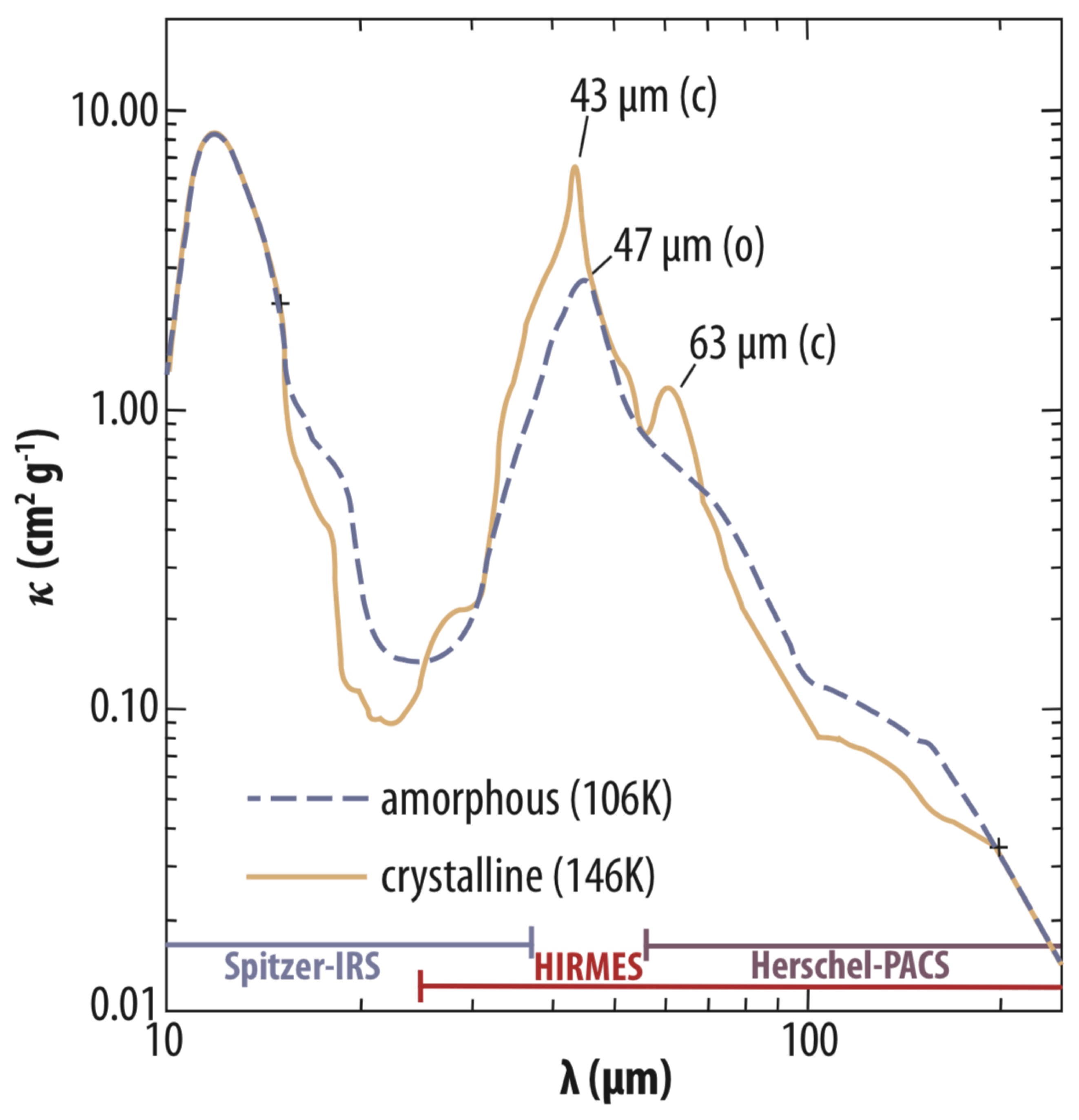}} 
\caption{An adaptation of \citet{2015ApJ...799..162M}'s figure, showing the the emission/absorption coefficients of the 43, 47 and 63 $\mu$m water-ice features that observationally infer the thermal history of grain mantles.} 
\label{fig:w-i}
\end{figure}

The ice/rock ratio in the solar nebula is thought to have been significantly larger than unity beyond the snow line. That is, the solid mass reservoir was dominated by ice by factors of 2 or more. Consequently, most core-accretion models form giant planets beyond the snow line \citep[e.g.,][]{2008ApJ...685..584I, 2008ApJ...688L..99D}. Compared to silicates and other refractory materials, water ice also substantially increases the sticking probability in collisions between dust grains, catalyzing the first stage of planet formation by growing micron-sized dust grains to centimeter and meter-sized icy bodies \citep{2008ARA&A..46...21B}. Solar system comets are likely primordial tracers of the original ice/rock ratios, and indeed suggest that ice was a dominant solid mass reservoir in the Solar System. While ices are often observed via their mid-infrared bands in the 3 -- 20 $\mu$m range, these features cannot generally be used to measure bulk ice in disks, as dust hot enough to emit at these wavelengths no longer retain ice. The longer-wavelength phonon modes (43 -- 63 $\mu$m), on the other hand, are expected to be seen in emission in typical proto-planetary disks. The strongest feature of crystalline ice at 43 micron has not been accessible since ISO \citep[see Figure \ref{fig:w-i};][]{1998A&A...332L..25M}. 

Beyond spectroscopy of proto-planetary disks, HIRMES will be able to access a diverse set of fine structure lines, including [FeII] 25.99 $\mu$m, 35.35 $\mu$m, [SI] 25.25 $\mu$m, [SIII] 33.48 $\mu$m, [SiII] 34.81 $\mu$m, [NeIII] 36.0 $\mu$m, [NIII] 57.30 $\mu$m, [NII] 121.90 $\mu$m, [OI] 63.18 $\mu$m, and [OIII] 51.81 $\mu$m, 88.35 $\mu$m. These, in particular, yield line ratios that probe the abundances, ionization state, and density in shocked gas. This will permit shock models to be tested, yielding quantitative mass flow rate measurements, as well as generating velocity-resolved line profiles that will elucidate the kinematics of shocked gas. Spitzer and Herschel have observed and mapped several such transitions in proto-stellar outflows and supernova remnants, allowing the spatial distributions of the various shock tracers to be compared. However, because their line profiles were unresolved, these observations did not supply kinematic information. HIRMES has the ability to provide the first complete data cubes in this field, allowing supersonic motions of the shock-heated gas to be measured as a key test of shock models. 

Designed to be as versatile as possible, HIRMES will be a major enhancement to SOFIA's suite of instruments, supplying the astronomical community with data not yet seen. We describe the instrument design in Section \ref{Section:Instrument}, and the various observing modes \& techniques and data reduction in Section \ref{Section:Observing}. The authors must stress that the design and sensitivities are still preliminary (correct at the time of writing). As the instrument undergoes Integration and Testing (I\&T) through Commissioning and Acceptance, it is likely that some parameters presented here will be updated. As such, we recommend the reader to keep visiting the SOFIA - HIRMES \href{https://www.sofia.usra.edu/science/instruments/hirmes}{webpage}\footnote{\href{https://www.sofia.usra.edu/science/instruments/hirmes}{https://www.sofia.usra.edu/science/instruments/hirmes}} for the most up-to-date information.


\section{HIRMES Instrument Design}\label{Section:Instrument}

\subsection{Overview}

The HIRMES instrument is a vacuum cryostat using a TransMIT pulse-tube cryocooler with a Commercial Off-The-Shelf (COTS) $^4$He refrigerator and an Adiabatic Demagnetization Refrigerator (ADR) to cool the Transition Edge Sensor (TES) bolometers to 70 mK. The instrument block diagram (see Figure \ref{fig:inst}; top) identifies the subsystems required to achieve the aforementioned science, including the cryostat, Fabry-Perot Interferometers (FPIs), optics, mechanisms, detectors, mechanical structures, and instrument control and data handling electronics. The following section sets out to detail the primary subsystems. A cut-section of the latest HIRMES CAD model is given in Figure \ref{fig:cut}. HIRMES is $\sim$1~m in diameter and $\sim$2~m in length. 

\subsection{Optical Design}

\subsubsection{Window}

HIRMES uses a 101.6 mm (4-inch) diameter (3-inch clear aperture) Topas\footnote{\href{https://topas.com/products/topas-coc-polymers}{https://topas.com/products/topas-coc-polymers}} (COC, Cyclic Olefin Copolymer) window, as the transmissive vacuum boundary between the instrument vacuum space and the Telescope Assembly (TA). Topas has low absorption over mid- to far-infrared wavebands and is also transparent in the visible, facilitating alignment and inspection. The average transmission is 93\%, with a reflection of 5\%, and an absorption of 2\%. A series of tests have been performed on the Topas window, including vacuum, stress, thermal, and environmentally testing a range of window thicknesses and curvatures. The result was to use a window with thickness = 400 $\mu$m and a radius of curvature = 66.5 mm. This radius of curvature is a formed curvature, concave to the instrument, which does not appreciably change under vacuum load. 

\subsubsection{Optics}

\begin{figure}[h!]  
\centering
\centerline{\includegraphics[width=0.9\linewidth]{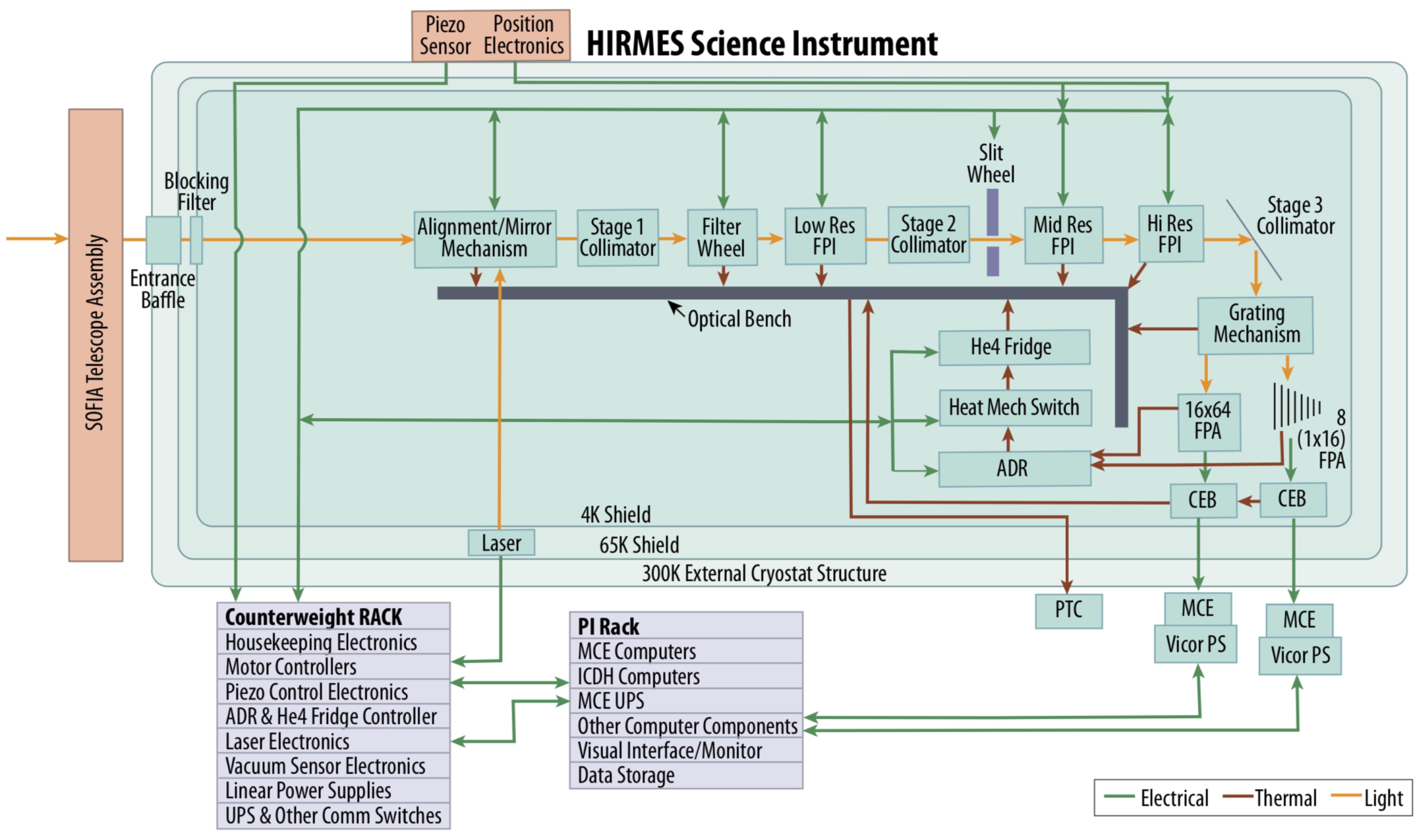}} 
\bigskip
\bigskip
\centerline{\includegraphics[width=0.9\linewidth]{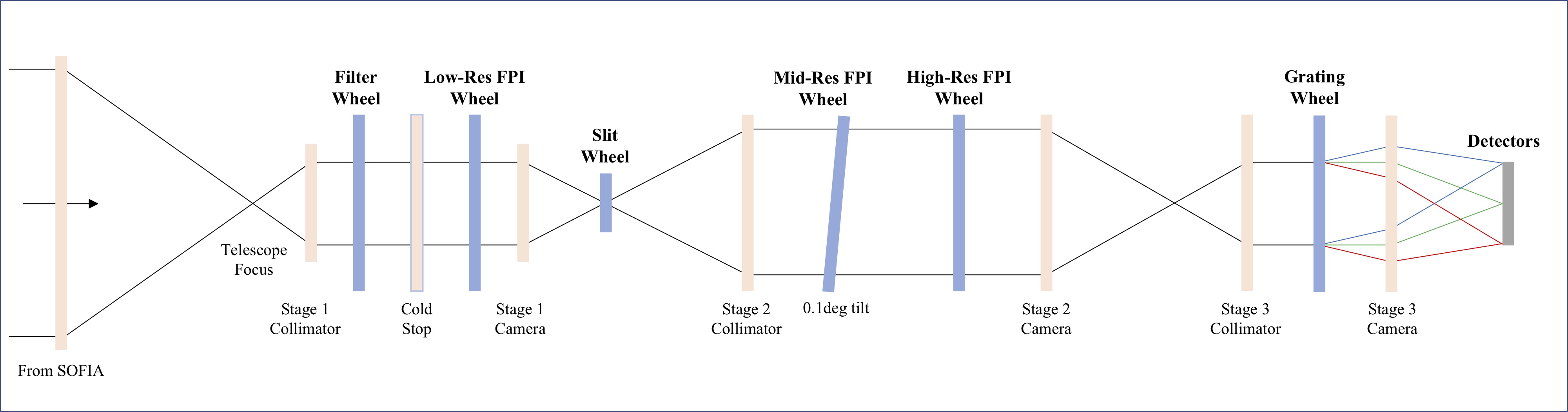}} 
\bigskip
\bigskip
\centerline{\includegraphics[width=0.9\linewidth]{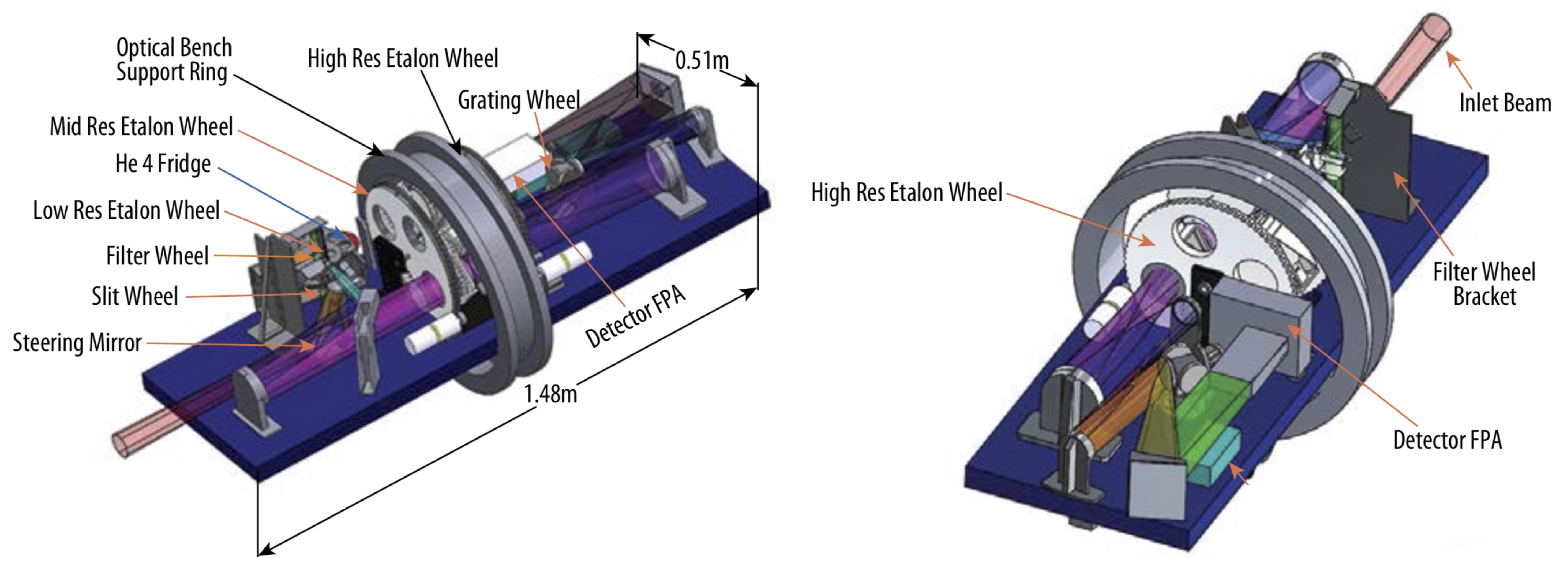}}
\caption{The HIRMES Sub-System Block Diagram shows the instrument’s internal and external components, the optical path with elements (middle, not to scale), and a labeled CAD model of the optical bench with opposing vantage points (bottom).} 
\label{fig:inst}
\end{figure}

The optical system provides efficient coupling to the telescope, controls stray light, provides filtering, and images the spectrum/image onto the detector arrays (see Figure \ref{fig:inst}; middle). The fixed optical components are manufactured with standard diamond turning process, and all of the components are fabricated from aluminum, thus the design of the optical bench is isothermal and contracts conformally. This allows the alignment to be completed warm with assurance that it will still be within tolerance when the instrument is cooled. The instrument opto-mechanical implementation is given in Figure \ref{fig:inst} (bottom).

\newpage 

\begin{figure}[h!]
\centering
\centerline{\includegraphics[width=0.81\linewidth]{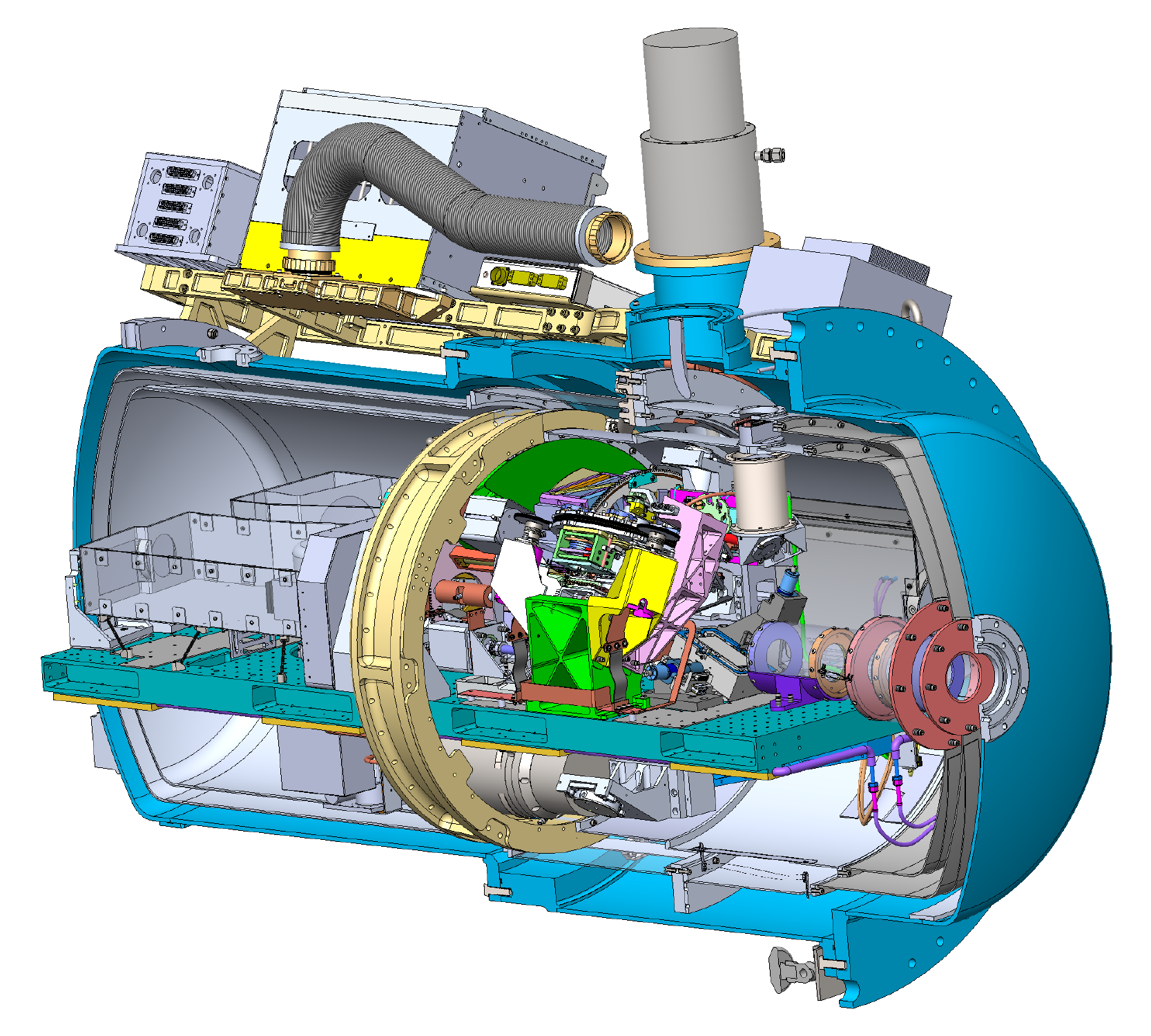}} 
\bigskip 
\caption{A cut-section of the HIRMES CAD model. Some components have been removed or phantomized to aid clarity. The light from the Telescope Assembly (TA) enters through the window on the right. \vspace{1cm}} 
\label{fig:cut}
\end{figure}

\smallskip
\noindent {\emph{Stage 1}:} 
\begin{addmargin}[1em]{0em}
The SOFIA telescope delivers a diffraction-limited f/19.5 beam over the HIRMES bandpass. HIRMES' slits are 113 arcsec long, with a selectable slit width from 3.0 to 8.7 arcsec to match the diffraction image size at the selected wavelength range. The slit length and the image field of view are well within SOFIA's 8~arcmin diameter field. The telescope's image rotation, generated by a moving observatory and the telescope's 3-axis spherical bearing mount, is handled operationally by adjusting the telescope's fine-drive to produce a fixed sky rotation for a period of time, typically 10 -- 30~mins depending on observational geometry \citep[for more details see][]{2014ApJS..212...24T}. As such there is no hardware image de-rotator. A system of two folding mirrors redirects the beam onto a collimator that produces a parallel beam at Stage 1. This Off-Axis-Paraboloid (OAP) collimator also forms an image of the telescope pupil of 20 mm diameter onto a cold stop. The filter wheel, accommodating 12 positions for the instrument filters, and the Low-Resolution FPI wheel, are placed at the approximate location of this pupil image. Following the Low-Resolution FPI wheel, there is the slit wheel with four slits and the open aperture for the Spectral Imaging mode. After the slit wheel, two folding mirrors redirect the beam onto the Stage 2 collimator. 
\end{addmargin}

\newpage
\newpage

\smallskip
\noindent {\emph{Stage 2}:} 
\begin{addmargin}[1em]{0em}
An OAP collimator produces parallel beams to go through the Mid-Resolution and High-Resolution FPIs filter wheels. It also produces a 80 mm diameter pupil image onto the Mid-Resolution FPI. The Mid-Resolution FPIs wheel is tilted by $\sim$0.1~degrees, and the High-Resolution FPIs are placed at 50~mm from the Mid-Resolution FPIs to eliminate parasitic ringing between the Mid-Resolution and High-Resolution FPI mirrors. Another OAP relays the telescope image onto another intermediate focal plane with an aperture to prevent scattered light going through the system. Due to the walk-off of the beam through the High-Resolution FPIs (primarily at the longer wavelengths) the outgoing beam is expanded.
\end{addmargin}

\noindent {\emph{Stage 3}:} 
\begin{addmargin}[1em]{0em}
Another OAP collimates the beam after the intermediate focus and forms a 40 mm pupil for the grating wheel, and the camera OAP relays the spectrally-dispersed light to the detectors. The flat mirrors in Stage 3 were added to package the optics into the available cold space. Final order-sorting is performed by a set of three reflective diffraction gratings. The linear dispersion at the detector array is set to spread the Free Spectral Range (FSR) of the Mid-Resolution FPIs onto at least 1 pixel at short wavelengths, and to spread the FSR wider than the point spread function (PSF) at long wavelengths. A rotary stage mechanism is used to select the grating and to set its angle over a range of $\pm$8.1\,degrees to choose the spectral line of interest. Each grating is blazed to optimize the efficiency over three sub-bands of the wavelength range. A mirror used for the spectral imaging mode occupies the fourth position on the stage. The spectral image (R$\sim$2000) is projected onto one side of the Low-Resolution detector array. 
\end{addmargin}

\medskip 
\noindent The optical imaging system is diffraction limited ($\ll \lambda/14$ at 24 $\mu m$) over the whole spectral range of the instrument. In fact, the quality of the reflective optical elements are such that they could be used for optical wavelengths. The instrument's internal stray light baffles are blackened with a ~500 micron thick layer of an absorptive mixture comprised of 65\% Epotek 377, 30\% fumed (pyrogenic) silica powder, and 5\% graphene by volume (alternatively, 50.7\% Epotek 377, 42.8\% fumed silica, and 6.5\% graphene by weight) \citep{2017RScI...88j4501C}. A monolayer of K1 borosilicate glass microspheres sieved to diameter $<$100 $\mu$m can be incorporated on this lossy dielectric surface and overpainted with $\sim$50 $\mu$m of Aeroglaze Z306\footnote{Lord Corporation Chemical Products, ``Aeroglaze Z306 Flat Black Absorptive Polyurethane Low-Outgassing Paint,'' 2000 West Grandview Blvd., P.O. Box 10038, Erie, PA 16514-0038}, where dilution through scattering is desirable to control the optical response. The resulting coating is CTE-matched (Coefficient of Thermal Expansion) for use on metallic substrates, non-magnetic, and robust under thermal cycling. Any rejected light is expected to exit the optical system or be absorbed by the partitioning blackened baffles in the front end of the instrument. 

\vspace{-0.2cm} 
\subsubsection{Filters}

The filters use a combination of technologies to cover the full 25 -- 122 $\mu$m spectral range: a dielectric multilayer filter to provide the 25 $\mu$m cut-on at short wavelength, and crystal filters at the longer wavelengths. Table \ref{tab:filters} details the specifications of each available filter mounted on the Stage 1 filter wheel (see Figure \ref{fig:inst}). The five ``First order FPI + long pass'' filters are used in the Spectral Imaging mode as order-sorting filters that each consist of a coupled fixed-width FPI and long-pass filter. Either plastic film or vapor deposited parylene are used for the anti-reflection coatings.

\begin{table}
\caption {Properties of the filters used in the Stage 1 filter wheel. \vspace{-0.25cm}} \label{tab:filters}
\begin{center}
\resizebox{\textwidth}{!}{\begin{tabular}{ |c|c|c|c| } 
 \hline
 Filter & Design & Details & Mean Transmission \\ 
 \hline
 23-32 $\mu$m; Bandpass & Multi-layer dielectric & On CdTe substrate & 70\% \\ 
 26 $\mu$m; Long Pass & Al$_2$O$_3$ with near IR blockers & Parylene AR (P.AR) coating & 75\% \\ 
 40 $\mu$m; Long Pass & Al$_2$O$_3$ + CaF$_2$ stack & P.AR coating on outer layers & 75\%  \\ 
 65 $\mu$m; Long Pass & Al$_2$O$_3$ + CaF$_2$ + BaF$_2$ stack & P.AR coating on outer layers & 75\% \\ 
 51.8 $\mu$m; [OIII] & First order FPI + long pass & Metal mesh & 50\% \\ 
 57.3 $\mu$m; [NIII] & First order FPI + long pass & Metal mesh & 50\% \\ 
 63.2 $\mu$m; [OI] & First order FPI + long pass & Metal mesh & 50\% \\ 
 88.4 $\mu$m; [OIII] & First order FPI + long pass & Metal mesh & 50\% \\ 
 121.9 $\mu$m; [NII] & First order FPI + long pass & Metal mesh & 50\% \\  
 \hline 
\end{tabular}} 
\end{center} \vspace{-3.5mm}
\end{table}

\newpage 
\begin{figure}[h!] 
  \centering
  \begin{minipage}[]{0.48\linewidth}
    \centering
    \vspace{-0.08cm}
    \includegraphics[width=0.9\linewidth]{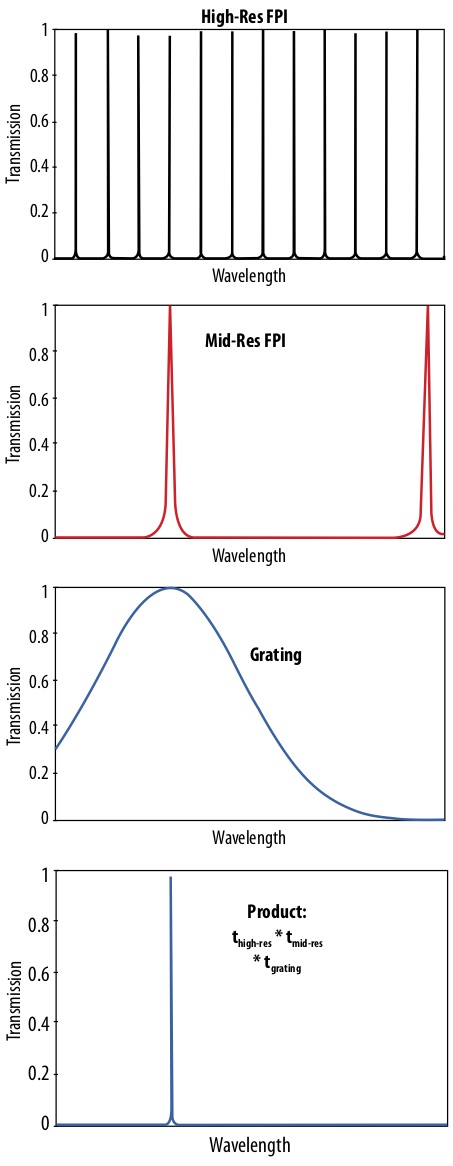}
    \caption{Top to bottom: transmission of high-res FPI, mid-res-FPI, grating, and their product. The product is very spectrally pure. These figures are purely graphical, and are not based on measurements.}
    \label{fig:fpi_comb}
  \end{minipage}
  \hfill
  \begin{minipage}[]{0.49 \linewidth}
    \centering
    \includegraphics[width=0.9\linewidth]{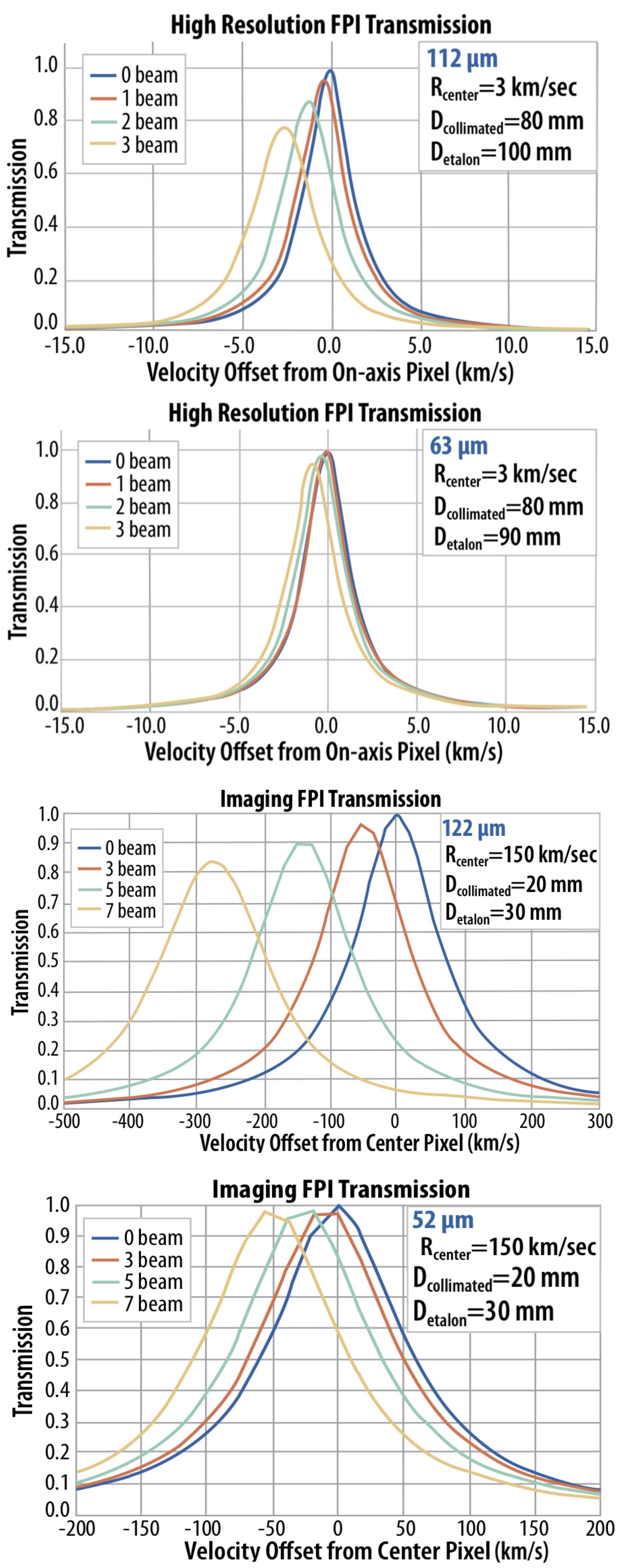}
    \caption{Top to bottom: transmission profile at 112~$\mu$m \& 63~$\mu$m (High-Resolution FPI) and 122~$\mu$m \& 52~$\mu$m (Imaging, Low-Resolution FPI) for beams on and off the optical axis. 1~beam~=~1~diffraction limited PSF at the detector.}
    \label{fig:fpi_off}
  \end{minipage}
  \vspace{1cm} 
\end{figure} 
\newpage

\vspace{-0.2cm} 
\subsubsection{Fabry-Perot Interferometers}

HIRMES uses a fleet of Fabry-Perot Interferometers (FPIs) for its various observing modes. The high spectral resolution mode of $R =$ 100,000 is achieved with FPIs. Also, FPIs with low spectral resolution ($R\sim$ 2000) will be used for Spectral Imaging.

An FPI consists of two highly reflective plane-parallel mirrors that form a resonant cavity. The resonance condition is $2 \cdot d = n \cdot \lambda$, where $d$ is the cavity spacing, $\lambda$ is the wavelength, and $n$ is an integer order. The spectral resolution of an FPI is the product of the finesse, $F$ (the FPI efficiency determined by the absorption in the reflector and the mean number of reflections in the cavity), and the order, $n$. To create a spectrum, the cavity spacing $d$ is adjusted, changing the resonant wavelength. Since any wavelength that fulfills the resonance condition will be transmitted by the FPI, it is necessary to employ additional filters to sort out the unwanted wavelengths. HIRMES uses additional FPIs with a resolution of about 12,000 (mid-resolution FPIs) and the reflective grating to select the desired wavelength (see Figure \ref{fig:fpi_comb}). 

Rays within either the axial beam or off-axis beams that pass the FPI at an angle with respect to the normal axis will resonate at a slightly shorter wavelength. The angle is minimized when the FPIs are located at pupil positions and have a large aperture, as is the case for the HIRMES FPIs. In addition, a high spectral resolution requires operating the FPI at a high order, hence a large cavity spacing. These two conditions determine the physical size of the FPIs. Although off-axis beams that pass the FPI at an angle with respect to the normal axis of the mirrors resonate at shorter wavelength, we will use this feature as an advantage. By slewing the telescope during the observation, and thus moving the science target from axial to off-axis pixels, we also spectrally sweep over a part of the nearby spectrum. To obtain the full desired spectrum thus requires less steps of changing the cavity spacing of the FPIs (see Figure \ref{fig:fpi_off}).

The FPIs in HIRMES use free-standing nickel metal meshes with a gold-flash as their highly reflective mirrors \citep{1963ITMTT..11..363U}. The finesse obtained with the metal meshes is a function of wavelength. Optimal finesses of FPIs made with these meshes ranges between 30-60 (a finesse that is too high would reduce the overall transmission). Thus, for each of the Mid- and High-resolution modes, HIRMES has three separate FPIs to cover the full wavelength range between 25 and 122 $\mu$m. The low-, mid-, and high-resolution FPIs have a scanning mechanism using piezo elements to select the desired cavity spacing, and step over a wavelength range to create spectra. In addition to the tunable FPIs, HIRMES will employ FPIs with a fixed cavity spacing that are tuned to specific fine-structure lines. The overall transmission that each of the FPIs will achieve is over 70\%. A full write-up of the FPIs used in HIRMES can be found at \citet{2018spie-arXiv180805218D} \& \citet{2018spie-arXiv180706019C}.

\begin{figure}
\centering
\centerline{\includegraphics[width=0.52\linewidth]{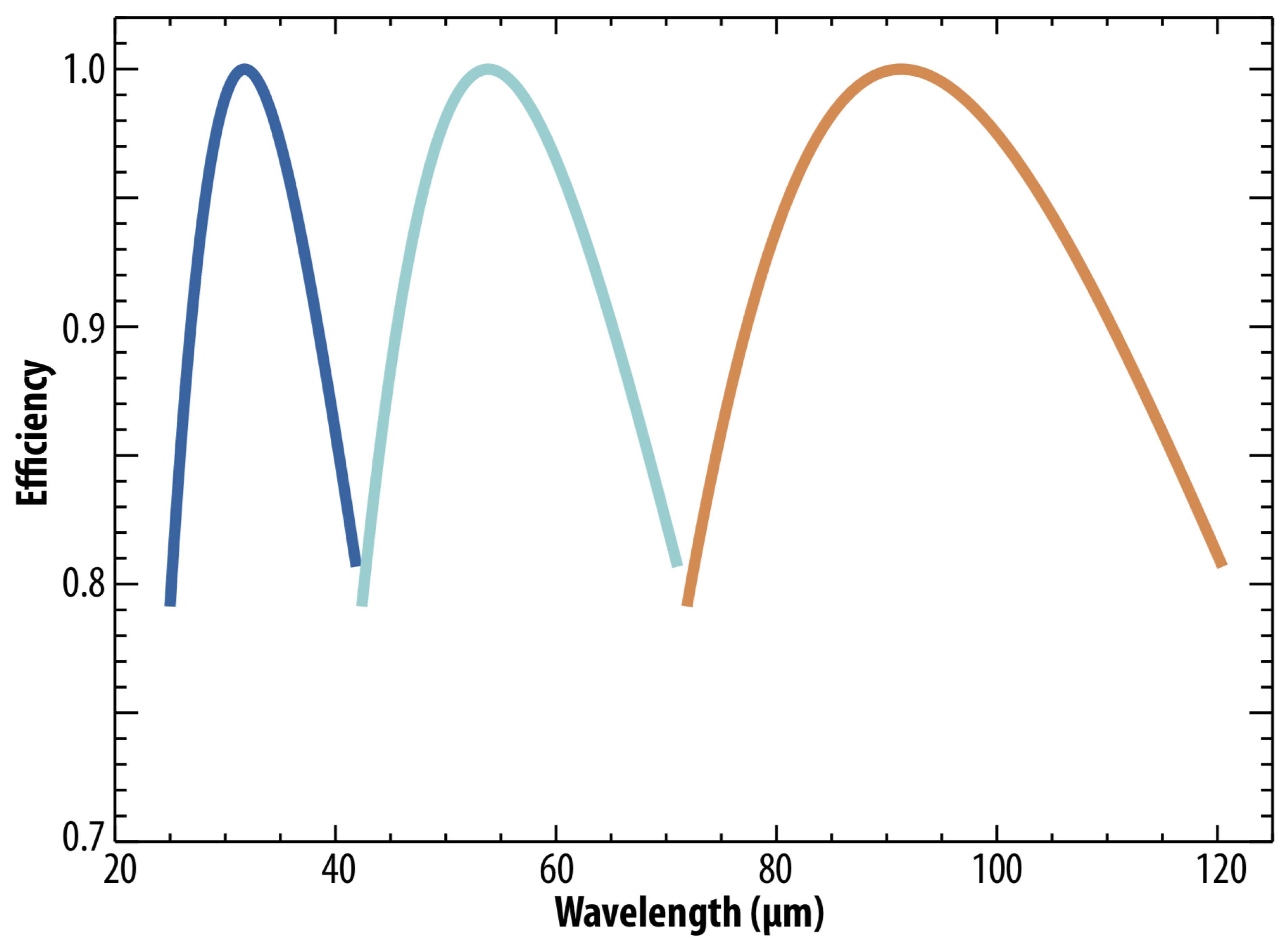}} 
\caption{The three selectable diffraction gratings provide $>$90\% average efficiency of both s- and p-polarizations over the HIRMES spectral region.} 
\label{fig:grating}
\end{figure} 

\subsubsection{Slits}

The slit wheel holds four slits of length = 113 arcsec, and widths = 8.7, 6.1, 4.2 and 3.0 arcsec respectively. These are selected based on the desired central wavelength and resolution. Additional to the slits, there is also a 2D image-stop of dimensions 113.0$\times$106.8 arcsec used in the Spectral Imaging mode, which is then projected onto 16$\times$16 pixels of the Low-Resolution detector array. 

\subsubsection{Gratings}
In the HIRMES wavelength range, echelette (blazed) gratings have near-ideal performance, so their efficiency can be calculated accurately using diffraction models and groove geometry. These gratings, with an the average efficiency of $\sim$0.9 (see Figure \ref{fig:grating}), were chosen to optimize performance at the most important spectral lines.  

\subsection{Detectors}
\begin{figure}
\centering
\centerline{\includegraphics[width=0.96\linewidth]{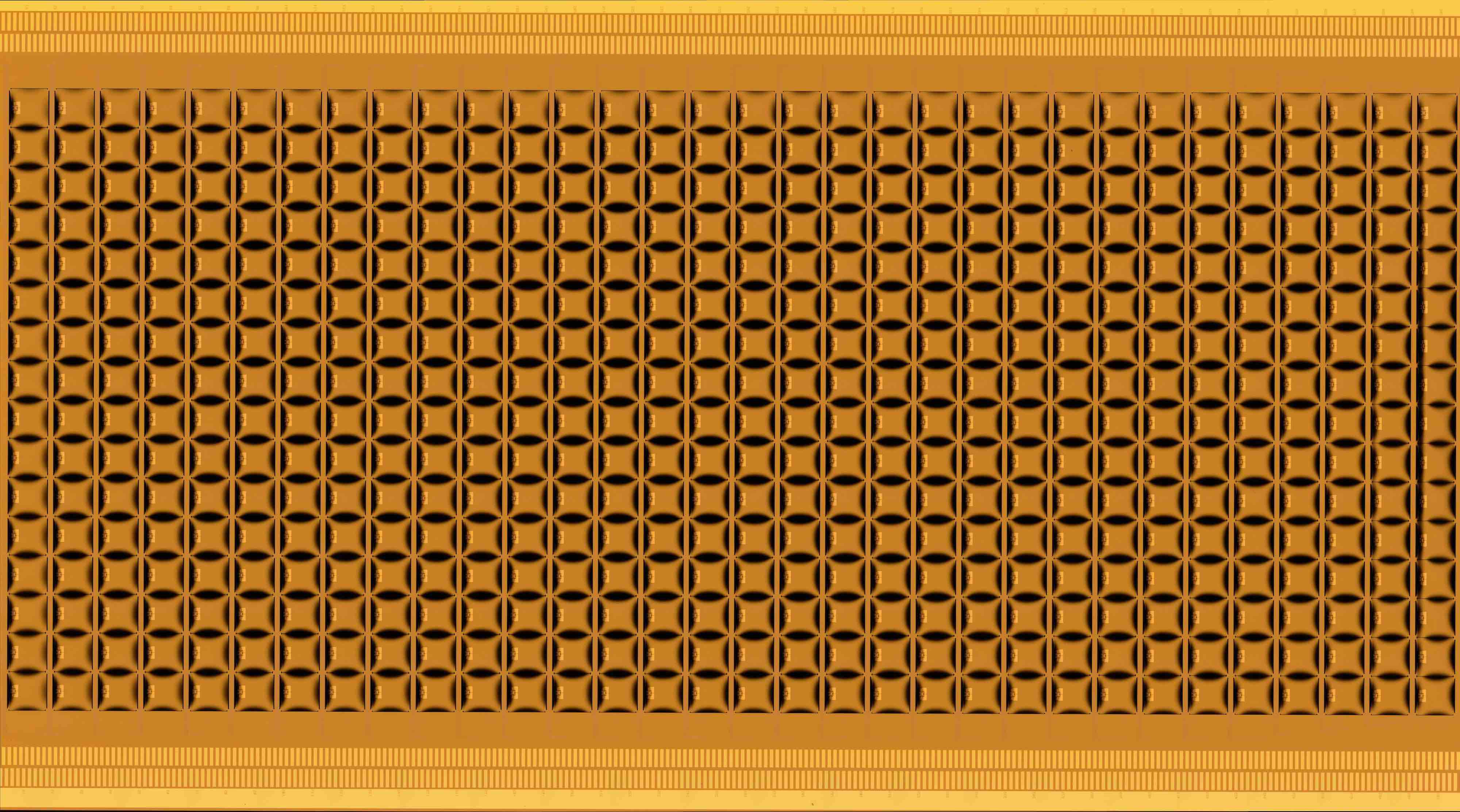}}
\vspace{0.1cm}
\centerline{\includegraphics[height=6.26cm]{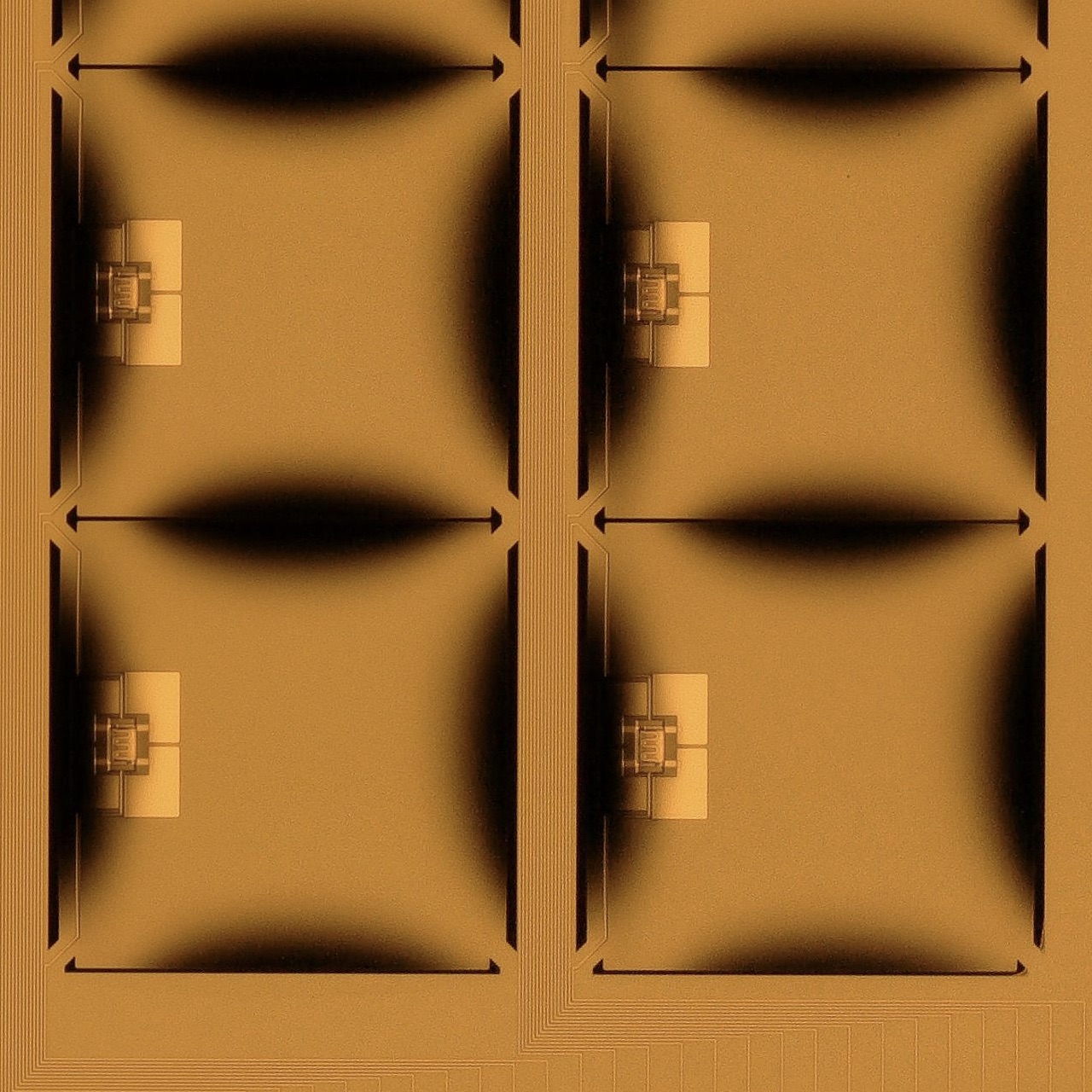}
\includegraphics[height=6.26cm]{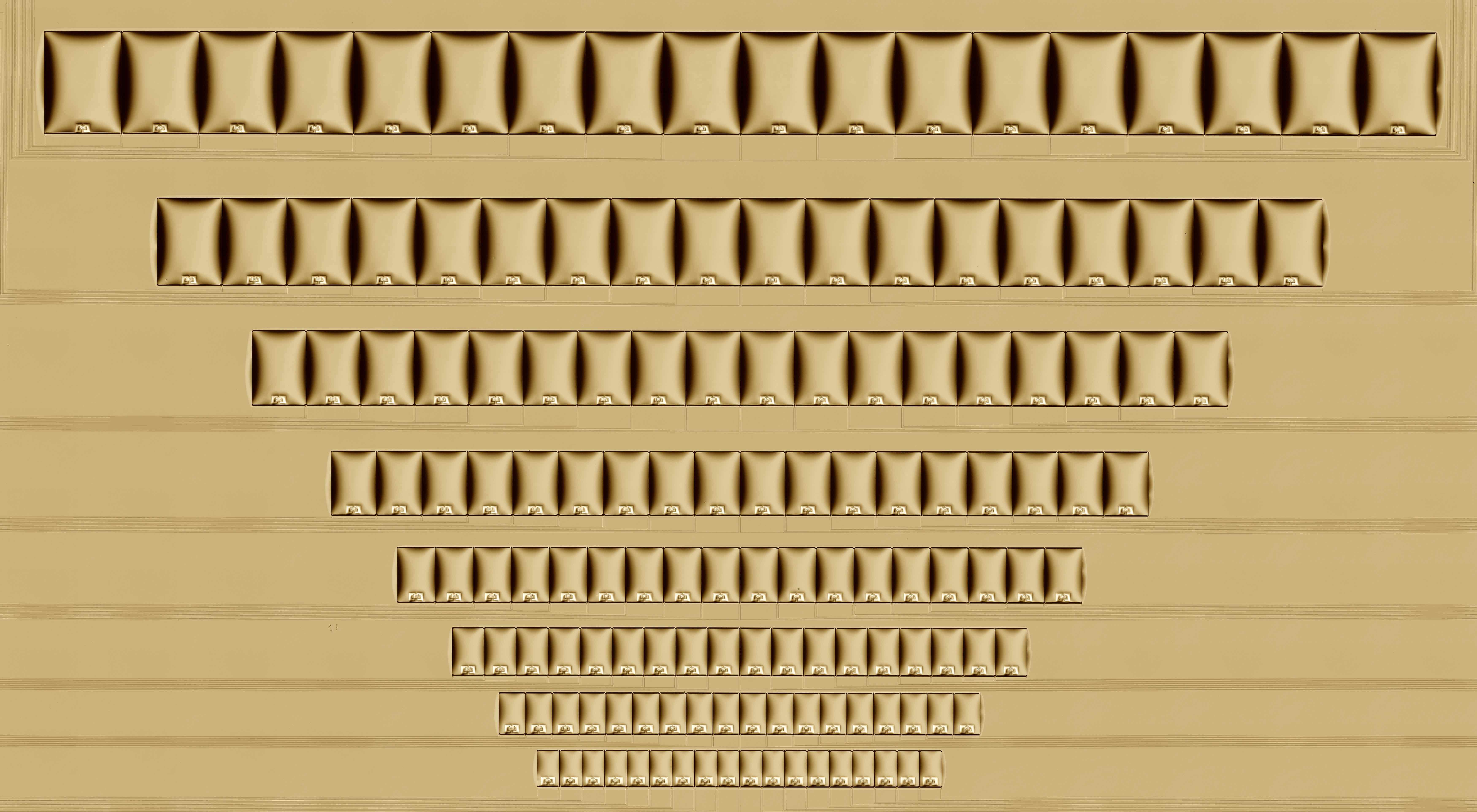}} 
\centerline{\includegraphics[width=0.75\linewidth]{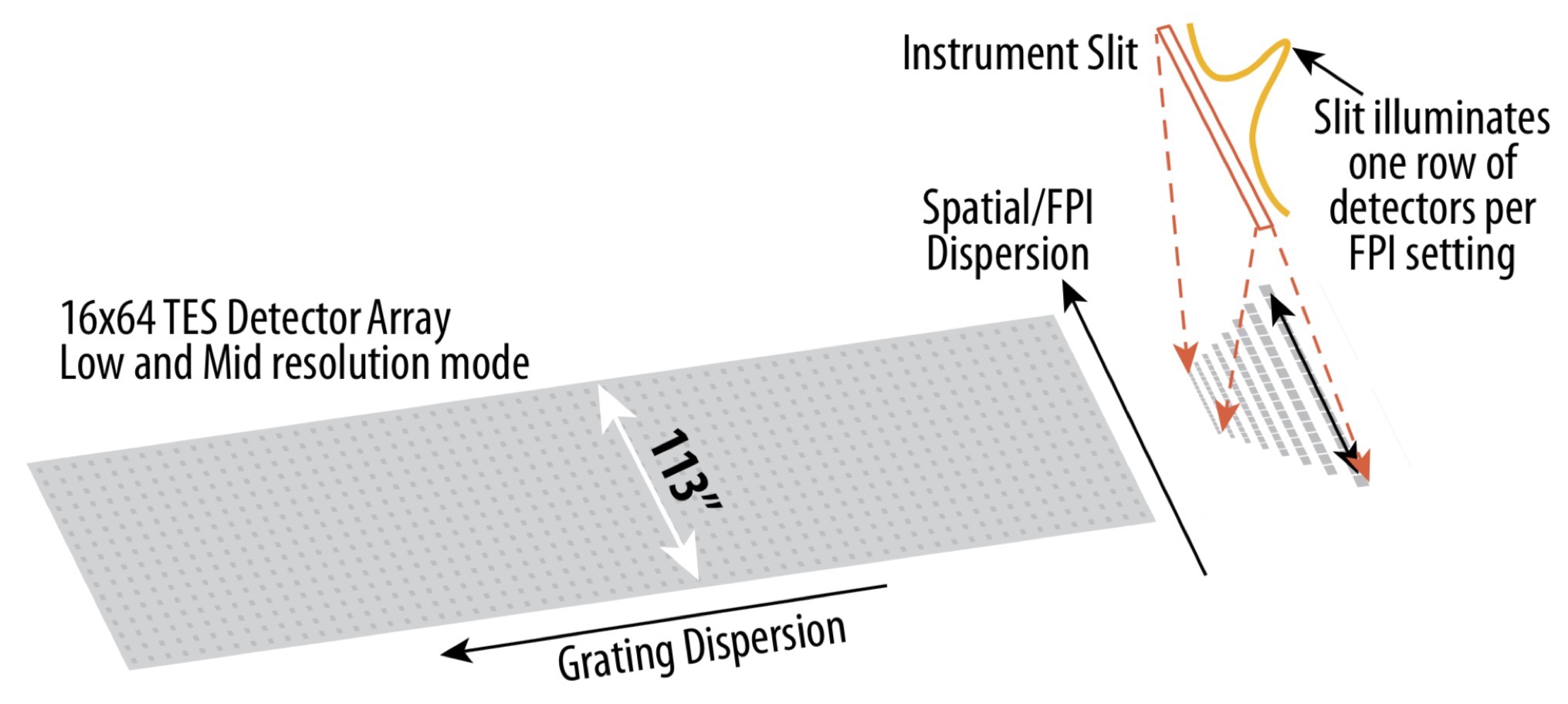}}
\caption{(Top) One half of the 64$\times$16 pixel Low-Resolution detector. (Middle-Left) A zoom in of the 4 most lower-right pixels of the Top image. (Middle-right) The eight, 1$\times$18 pixel subarrays of the High-Resolution detector, in which the pixels located at the edges of each subarray are not read out, resulting in 1$\times$16 active pixel subarrays. (Bottom) A schematic of the layout of both detectors on the Focal Plane.} 
\label{fig:detectors}
\end{figure}

The HIRMES detector layout consists of two Transition Edge Sensor (TES) bolometric detector arrays (see Figure \ref{fig:detectors}) that equip HIRMES with eight subarrays of 1$\times$16 pixels for the High-Resolution mode (low saturation power), and a 64$\times$16 array for the Mid- \& Low-Resolution and Spectral Imaging modes (high saturation power). A full description of the HIRMES detectors can be found in \citet{2018JLTP..tmp...89B} \& \citet{,2018JLTP..tmp..146B}. 

The ``Low-Resolution'' 64$\times$16 array is comprised of 1 mm $\times$ 1 mm square pixels with a 50\% absorptive frequency-independent coating. The thermal isolation design  consists of eight single-crystal silicon legs that are 1.4 $\mu$m thick, 50 $\mu$m wide, and 30 $\mu$m long, and will provide an expected Noise Equivalent Power (NEP) $\sim$2x10$^{-17}$ $W/\sqrt{Hz}$ and saturation power $\sim$25 pW, inclusive of the 50\% absorption efficiency. 

The ``High-Resolution'' array is comprised of eight, 1$\times$16 pixel subarrays, and is operated in a manner in which only one detector subarray is used at a given time. Each subarray is separately optimized for operation at a specific wavelength, by matching to the Full-Width Half-Maximum (FWHM) of the beam at the wavelength, and providing efficient absorption using a quarter-wave back short \citep{2018JLTP..tmp..116M}. The central wavelength for each subarray is: 30, 36, 43, 51, 61, 73, 88 \& 105~$\mu$m, and their respective physical pixel-widths range from 0.4 -- 1.4~mm. This system, with background-limited detectors, is near optimal for the High-Resolution spectroscopy mode throughout the entire spectral range. The detectors are Mo/Au TES bilayers deposited on photolithographically-defined leg-isolated 0.45 $\mu$m thick, 5 $\mu$m wide, and 30 $\mu$m long single-crystal silicon membranes. These detectors have an expected NEP $\sim$3x10$^{-18}$ $W/\sqrt{Hz}$ and saturation power $\sim$0.13 pW. 

The detectors are read out using NIST 1$\times$11 multiplexers operating at the detector temperature of 70 mK. The signal is amplified by a SQUID series array operating at $\sim$4 K, and controlled by a room temperature Multi-Channel Electronics (MCE) controller \citep{2013Hasselfield,2016SPIE.9914E..1GH}. The Low-Resolution and High-Resolution detectors are packaged in a single, common Focal Plane Assembly (FPA), and kinematically mounted using a Kevlar and \mbox{magic-Titanium [15-3-3-3]} support system to provide thermal and mechanical isolation.   

\subsection{Cryostat \& ADR}

HIRMES implements a tri-layer all aluminum (6061-T6) cryostat design (see Figures \ref{fig:cut} \& \ref{fig:inst}), with the load path extending from the flange ring (where the instrument mounts to the Telescope Assembly), through the central `bellyband' support region of the cryostat. The external hemispherical end caps, as well as the 65 K and 4 K heat shields, are supported from this bellyband, along with the monolithic 4 K optical bench (from a three-point mount). This bench supports all working instrument components and optics. The heat shields and optical bench are supported by aluminum rings suspended using 12 titanium struts with titanium end caps from the belly band. The optical bench, bulkheads, and structural components are made of the same forged 6061-T6 aluminum as the cryostat, providing high strength and thermal expansion coefficients that match the mirrors and mirror mounts. All aluminum mirrors and the optical bench are machined, annealed, and assembled to ensure dimensional stability at cryogenic operating temperatures, preventing hysteresis from repeated cool-downs.

The HIRMES thermal design uses a two-stage Pulse Tube Cryocooler (PTC) to cool the instrument and optical bench. A $^4$He sorption fridge coupled to an Adiabatic Demagnetization Refrigerator (ADR; salt-pill) is subsequently used to cool the detectors located in the FPA to 70 mK, with an expected operational hold-time of $>$12 hours. A TransMIT PTD406C pulse-tube head is mated to one of the two SOFIA on-board Cryomech CP2870 He compressors. The PTC first stage cools the outer radiation shield below 65 K. This lowers the radiation heat load from the cabin-temperature cryostat shell to a level that allows the second stage to cool the optical bench and inner radiation shield to 4 K. 

\newpage

A $^4$He sorption fridge then lowers the ADR heat sink temperature to 1.3 K, allowing a quicker recycle time and enhanced cooling power. From this low starting temperature, the ADR is able to cool the detector to 70 mK. Additionally, a separate $^3$He sorption fridge acts as a 0.3 K thermal intercept to the ADR, reducing its thermal load stress. 

\subsection{Instrument Calibration}

HIRMES calibration captures the wavelength settings for observations, instrumental spectral profile, and radiometric calibration to allow the line intensity to be determined. The spectral calibration uses gas cells for absolute wavelength calibration on the ground, and two Quantum Cascade Lasers (QCLs) for in-flight verification ($\sim$ 63 \& 83 $\mu$m, R $\sim10^6$). The QCLs are mounted to the 65 K stage, and are injected into the instrument by reflection off a mirror on the back of the slit wheel. The QCLs provide good short-term wavelength stability; absolute knowledge is provided by a fixed etalon used with the QCL. With knowledge of the spectral setting and instrumental profile, radiometric calibration is achieved by observing known continuum sources, including rocky moons and asteroids. During operation, the FPI spacing is measured and monitored with FPI capacitive sensors. Parallelism is established at room temperature and only requires minor adjustment when cold by a known, fixed amount, through the usage of low-voltage, tilt PICMA piezo actuators from Physik Instrumente. 


\section{Observing with HIRMES}\label{Section:Observing} 

\begin{table}
\renewcommand{\arraystretch}{1.25} 
\caption {Properties of the filters used in the Stage 1 filter wheel.} \label{tab:modes}
\begin{center}
\resizebox{\textwidth}{!}
{\begin{tabular}{ | l | c | c | c | c | } 
 \hline
 \multicolumn{1}{|c|}{Parameters} & High-Res & Mid-Res & Low-Res & Imaging \\ 
 \hline
 \hline 
 Sensitivity (5$\sigma$, 1hr) & $\lesssim 1 \times 10^{-17} W/m^2$ & \multicolumn{3}{c|}{$\sim1 \times 10^{-16} W/m^2$} \\ \hline
 Resolving Power (R = $\lambda/\delta\lambda$) & 50,000 -- 100,000 & $\sim$12,000 & 325 -- 635 & $\sim$2,000 \\ \hline
 Angular Resolution & \multicolumn{4}{c|}{Diffraction Limited} \\ \hline
 Slit Size / FOV (arcsec) & \multicolumn{3}{c|}{Length: 113''; Width: 8.7'', 6.1'', 4.2'' \& 3.0''} & 113.0'' $\times$ 106.8'' \\ \hline
 Spectral Range & \multicolumn{3}{c|}{25 -- 122 $\mu$m} & Selected lines$^A$ \\ \hline
 Simultaneous Spectral Coverage ($\delta\lambda/\lambda$) & $\lambda$/R & \multicolumn{2}{c|}{0.1$\lambda$} & 0.001$\lambda$ \\ \hline
 Detector Format & 8$\times$16 pix$^B$ & \multicolumn{3}{c|}{64$\times$16 array$^C$} \\ \hline
 Detector Type & \multicolumn{4}{c|}{Transition Edge Sensor (TES)} \\ 
\hline 
\end{tabular}}
\end{center}
\footnotesize
$^A$Single wavelength setting for selected filters (63.2 $\mu$m [OI]; 51.8 $\mu$m, 88.4 $\mu$m [OIII]; 57.3 $\mu$m [NIII]; 121.9 $\mu$m [NII]). \\ 
$^B$High resolution detector consists of eight, 1$\times$16 pixel linear subarrays; whose pixel size increases per subarray. Shorter \\
$^{\ \ }$wavelength light is positioned onto the smaller pixel subarrays, longer wavelength light onto the larger pixel subarrays. \\ 
$^C$Spectral Imaging uses only a 16$\times$16 pixel section of the 2D array.
\end{table}

\subsection{Observing Modes}

\begin{figure}[h!]
\centering
\centerline{
\includegraphics[width=0.495\linewidth]{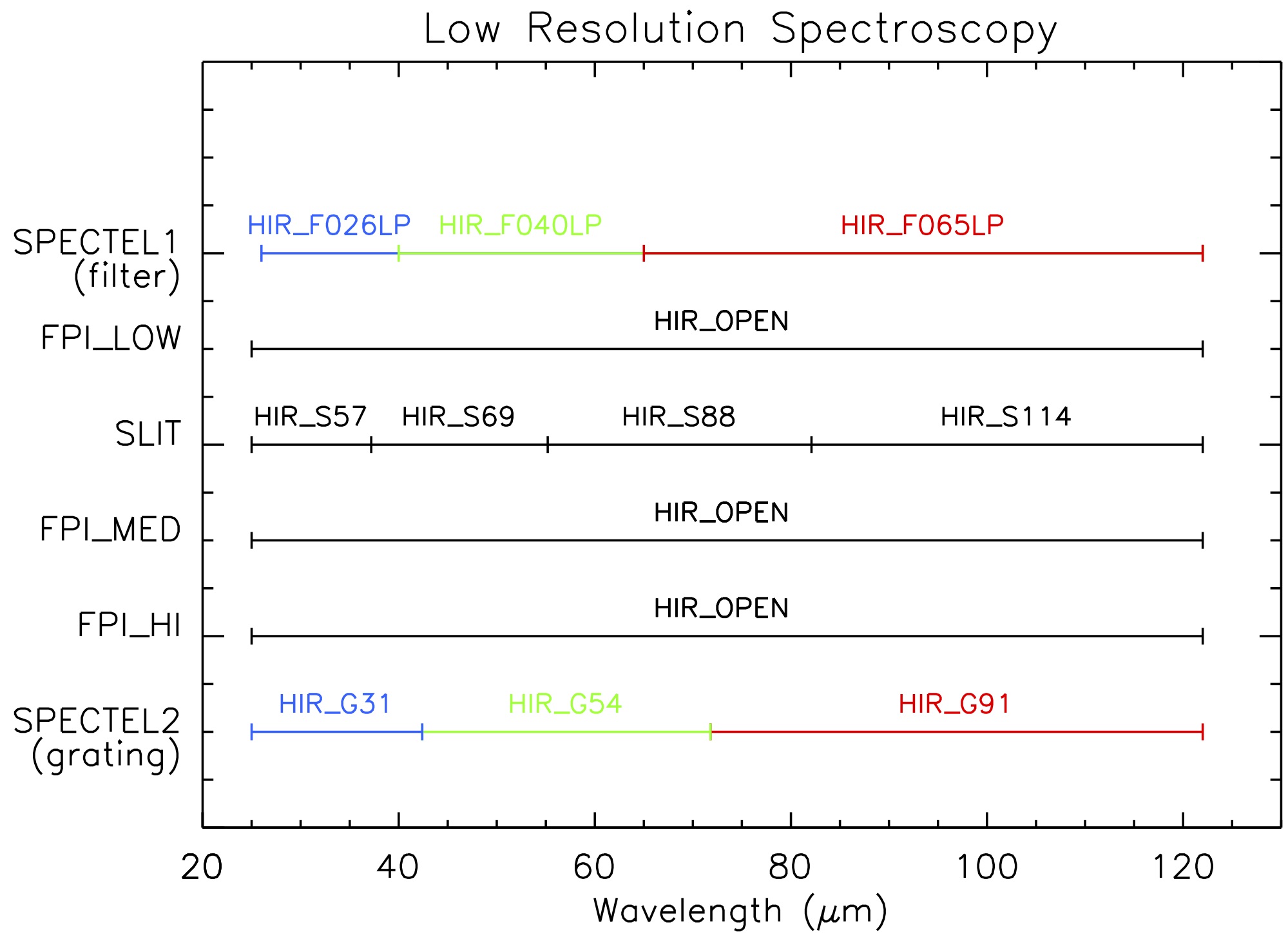}
\hfill
\includegraphics[width=0.495\linewidth]{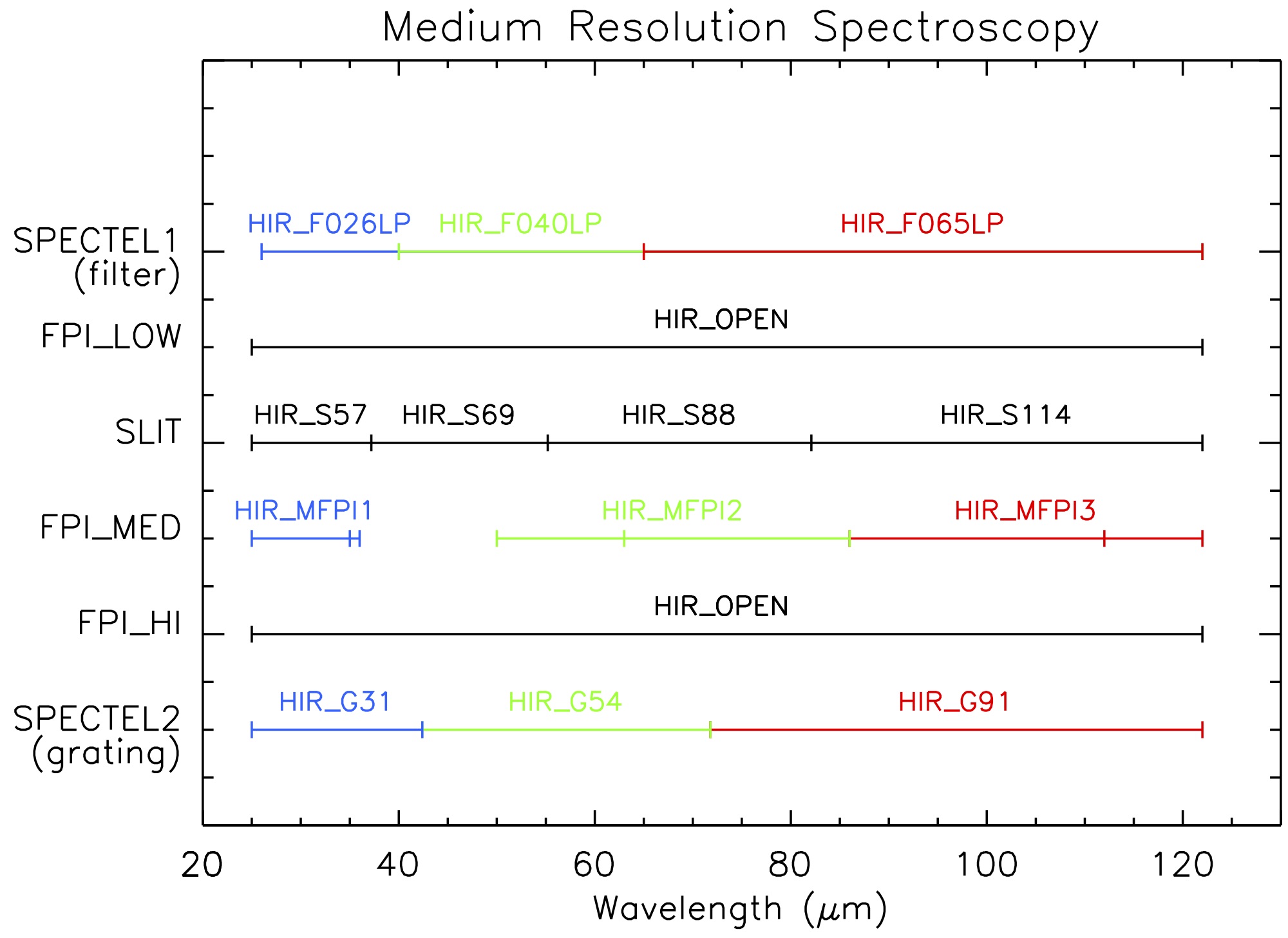}
} 
\bigskip 
\centerline{
\includegraphics[width=0.495\linewidth]{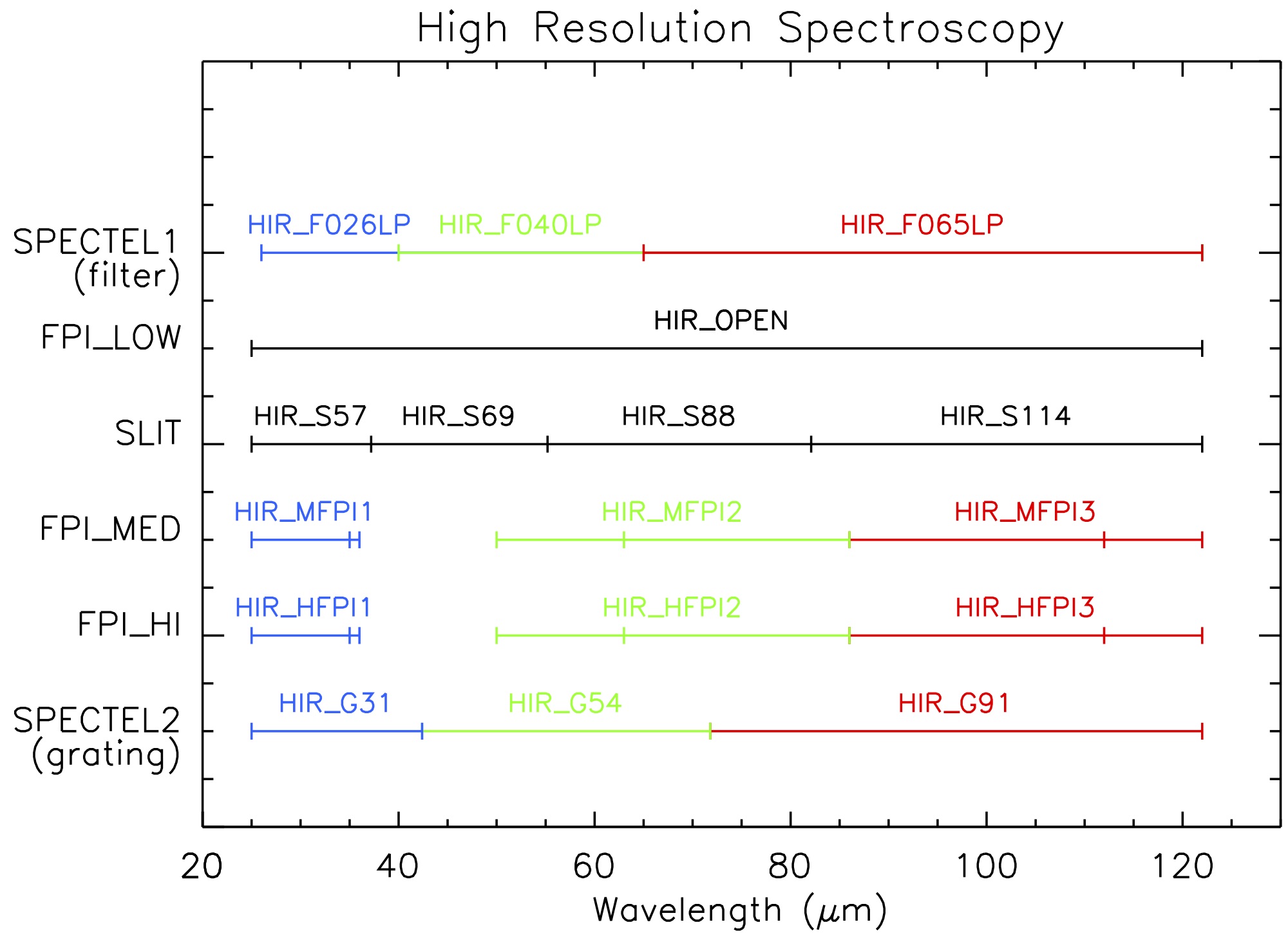}
\hfill
\includegraphics[width=0.495\linewidth]{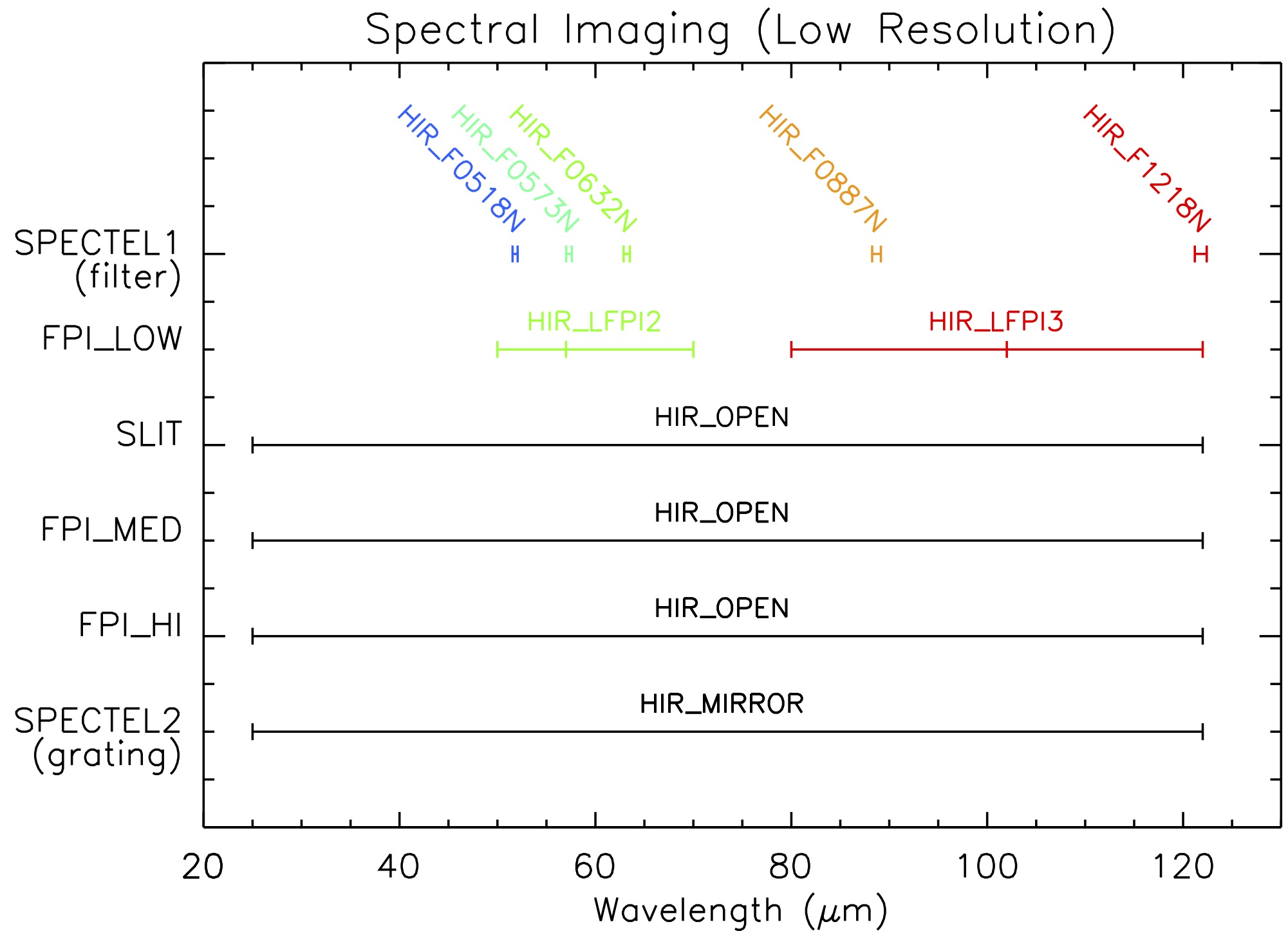}
}
\bigskip 
\centerline{
\includegraphics[width=0.495\linewidth]{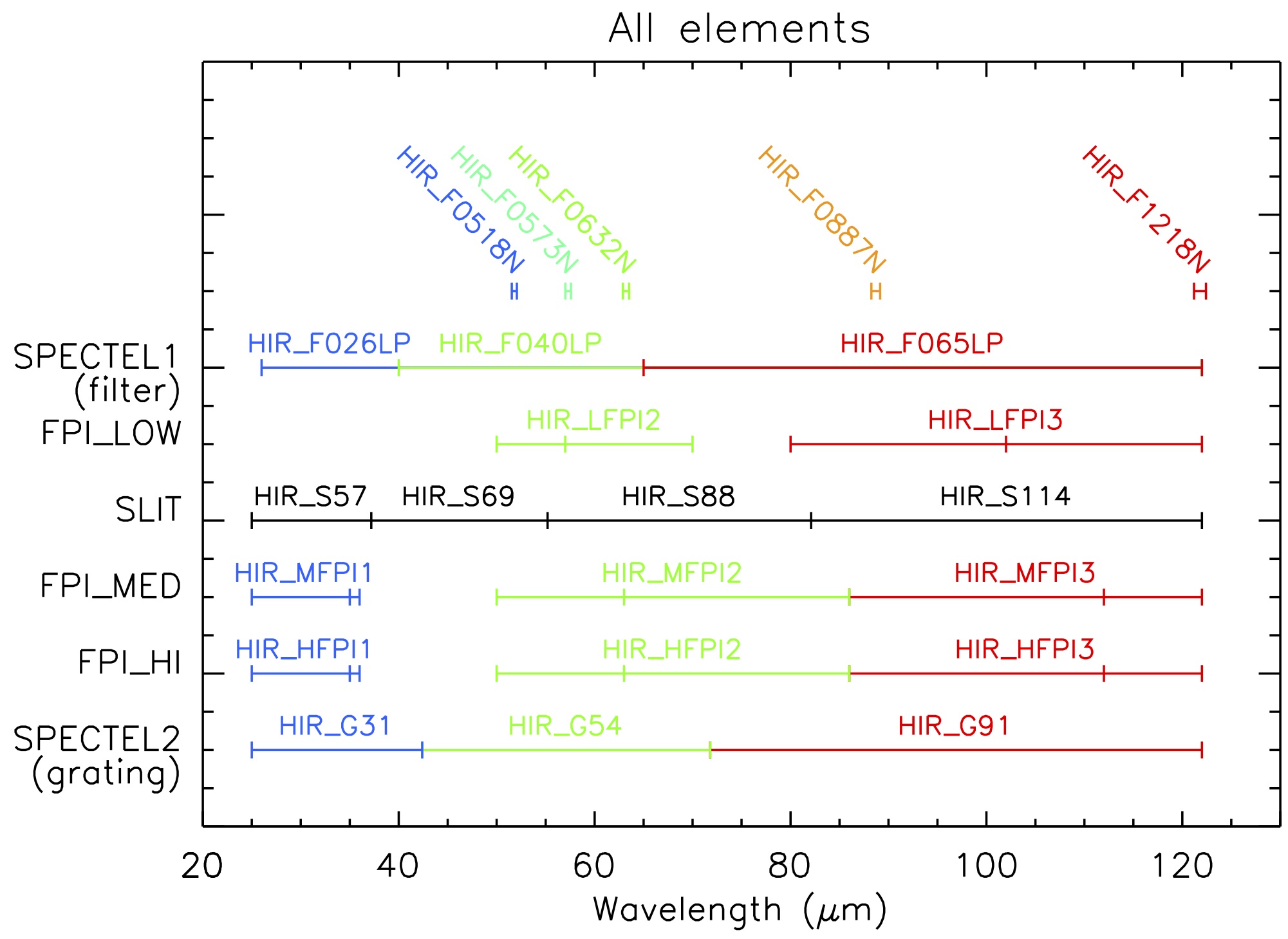}
\hspace{0.6cm}
\includegraphics[width=0.45\linewidth]{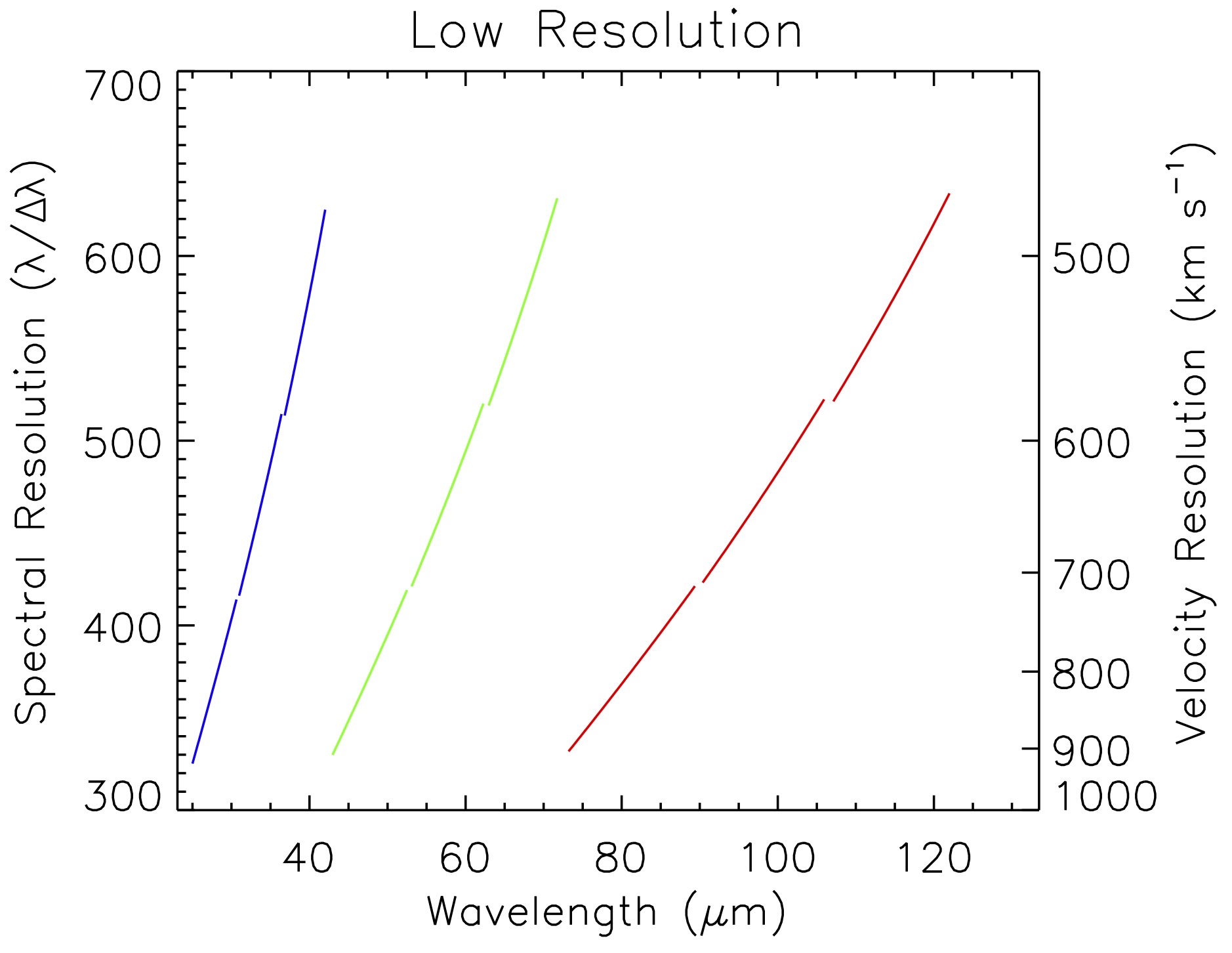}
}
\caption{The four primary observing modes and their combination of optical and spectral elements. The lower-left plot shows all optical \& spectral elements.  The lower-right plot shows the resolution as a function of wavelength for the Low-Resolution grating mode. Note that it would take nine different wavelength settings to obtain the full 25 -- 122 $\mu$m spectral range in this mode (three for each grating). } 
\label{fig:elements}
\end{figure}

By combining the direct-detection arrays (TES bolometers), grating-dispersive spectroscopy, and a host of Fabry-Perot tunable narrow-band filters, HIRMES can provide four primary observing modes: High Resolution (R$\sim$100,000), Mid-Resolution (R$\sim$10,000) and Low-Resolution (R$\sim$600) spectroscopy, and Spectral Imaging (R$\sim$2000). Figure \ref{fig:elements} shows how the various optical and spectral elements can combine to support each primary observing mode. Figure \ref{fig:fpi_comb} \& \ref{fig:fpi_off} are useful to refer to when reading through the following mode descriptions. HIRMES is a complex instrument with many configurable elements, however, the combination selection for various modes will be operationally automatic, based on the desired science. One only needs to select the central wavelength, delta wavelength, number of steps, and one of the four modes. If instrumental resolution is desired, then one only needs to specify the wavelength range and the mode. Having said that, any combination of optical and spectral elements is technically feasible to create additional modes. A summary table of the observing modes and their respective properties is given in Table \ref{tab:modes}. 

\newpage
\newpage 

\subsubsection{Low-Resolution Spectroscopy}
The light first passes through a bandpass filter, then a slit, then a reflective grating. The spectrum is under-sampled by the 64$\times$16 pixel detector, producing a resolution of R $\sim$ 320 -- 635 (see Figure \ref{fig:elements}, lower-right), and instantaneous bandwidths of 5 -- 15 $\mu$m, depending on the wavelength. It would take nine different wavelength settings to obtain the full 25 -- 122 $\mu$m spectral range in this mode.

\subsubsection{Mid-Resolution Spectroscopy}
This mode keeps the configuration of the Low-Resolution grating mode, however inserts a Mid-Resolution FPI into the optical path.  The effect of this is a narrow, sharp wavelength peak that is greatly under-sampled in one of the pixels of the 64$\times$16 pixel array. The FPI is stepped through one free spectral range (in roughly 10-50 steps depending on wavelength and desired spectral sampling) to produce a spectrum over the full instantaneous spectral coverage of the Low-Resolution grating mode. 

\subsubsection{High-Resolution Spectroscopy}
Going one step further to the Mid-Resolution mode, an additional High-Resolution FPI is inserted into the optical path, and the central wavelength is centered onto the appropriate column of the eight 1$\times$16~pixel linear subarrays of the High-Resolution detector (which linear subarray depends on wavelength; subarray pixel size is proportional to wavelength). That means there is only a single pixel in the spectral dimension, which is closely matched to the PSF. However, due to the radially dispersive nature of FPIs, stepping spatially up and down the slit also results in a slight wavelength shift ($\sim$ 1 -- 5 km/s depending on wavelength, see Figure \ref{fig:fpi_off}). Combining this feature with stepping the FPIs by discrete steps, results in the spectral sampling of a desired wavelength range (typically a single narrow spectral line). Figure \ref{fig:fpi_comb} visualizes the product of the FPIs working together with the grating to produce a High-Resolution line spectrum. 

\subsubsection{Spectral Imaging}
This mode changes the configuration completely by switching out the initial bandpass filter with a narrow-band filter on the same filter wheel. This narrow-band filter is actually a combination of a fixed-width FPI, and its own bandpass filter, both of which are tailored for specific spectral lines (see Table \ref{tab:filters}). The other configurable elements used in optical path are a Low-Resolution FPI, a square image-stop instead of a slit, and a mirror instead of the grating. The 2D image is then placed on one side of the 64$\times$16 pixel array, to produce a 16$\times$16 pixel spectral image (113.0$\times$106.8 arcsec), whose wavelength is also variable over the image, due to the radially dispersive nature of the FPI (see Figure \ref{fig:fpi_off}).  

\subsection{Sensitivity Limits}

\begin{figure}
\centering
\centerline{\includegraphics[width=0.8\linewidth]{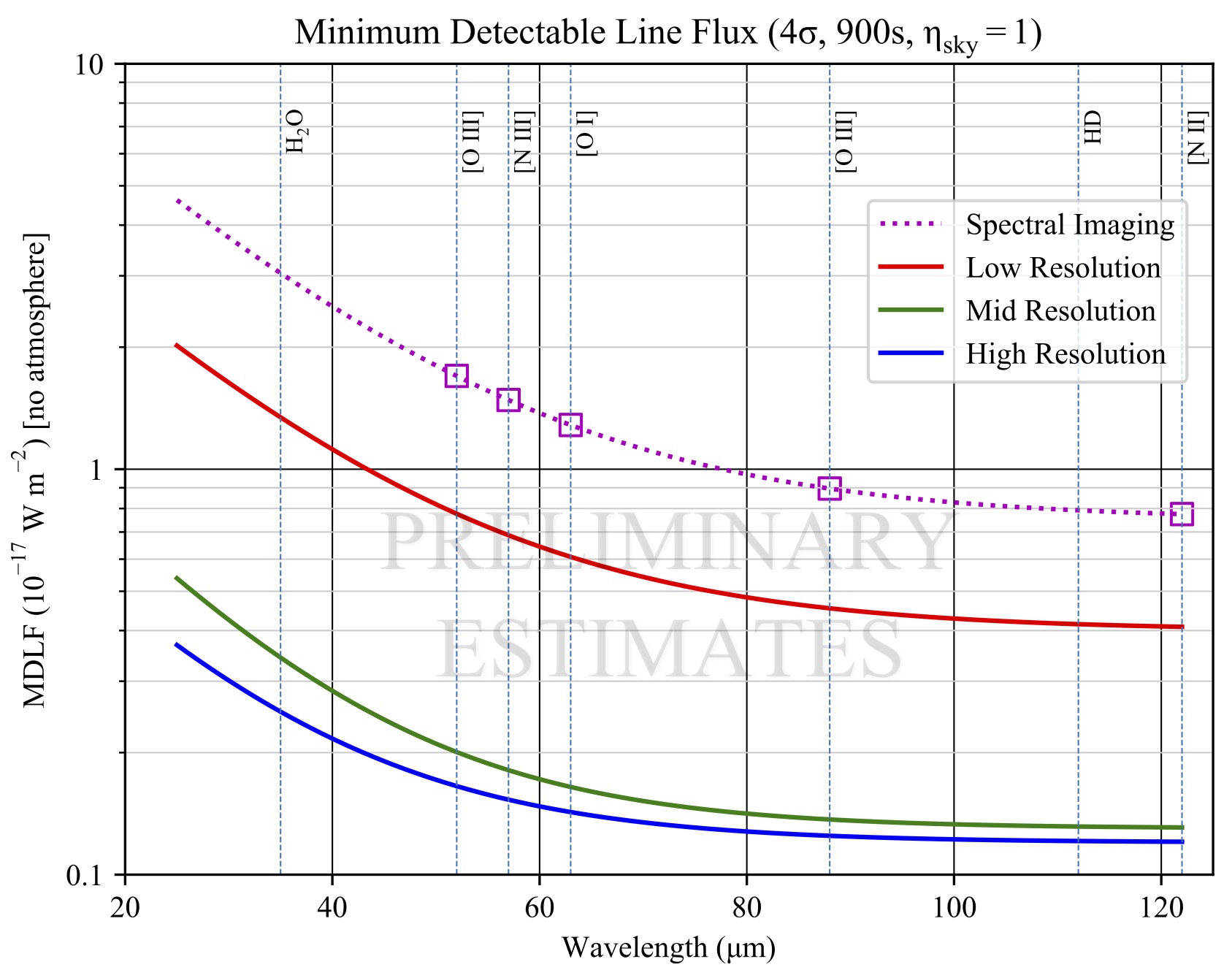}} 
\caption{Visual representation of the Minimum Detectable Line Fluxes (MDLFs) for the different modes, assuming no atmosphere. Note that the Spectral Imaging mode cannot cover the entire wavelength range, only the discrete wavelengths, shown by square line-markers. The wavelengths of key spectral line features are marked by vertical blue dotted lines.}  
\label{fig:sensitivity}
\end{figure}

As the instrument is still being built, full characterization and calibration of the various components has yet to be completed. The instrument sensitivities or capabilities presented here are the best current estimates based on analysis of the design. In particular, Figure \ref{fig:sensitivity} visualizes the estimates of Minimum Detectable Line Fluxes (MDLFs) for the various observing modes. One can extrapolate an estimate of the total time on-source for a given flux from Figure \ref{fig:sensitivity}, and later apply atmospheric transmission factors. Observational overheads are not taken into account, and will be fully realized once cold function checks and commissioning are underway. 

\subsection{Observing Techniques }

Each detector read-out has associated astrometric and timing data, enabling HIRMES to be able to support any TA Observing mode (e.g. Lissajous, raster-scan, slit-scan, mapping, chop \& nod, etc.). The typical TA Observing mode when in either of the Low-, Mid-, or High-Resolution modes would be slit-scan (scanning up and down the length of the slit), and Lissajous for Spectral Imaging. Both of these modes would be performed without chopping.

The intent of selecting slit-scan and Lissajous for typical TA Observing modes is two-fold: 1) to maximize the spectral bandwidth by moving spatially across the FPIs, and 2) to break the degeneracy of having the same sky and/or source flux on the same pixels, thereby increasing the ability to characterize the detector for data reduction. This is also achieved by a new TA Observing mode in development that allows the sky to rotate whilst tracking. Atmosphere-less moons and asteroids will be regularly observed, in addition to blank-sky / sky-dips, for flux calibration and telluric correction. 

\newpage 

\subsection{Control and Data Reduction Software}

The HIRMES software system design includes: 1) Graphical User Interface (GUI) for instrument control and monitoring; 2) detector data acquisition; 3) detector and readout tuning and control; 4) calibrate mechanisms and spectroscopic elements; 5) interface with ancillary devices; 6) interfacing with telescope and executing observations via the SOFIA Command Language (SCL); 7) power and thermal management; 8) health and safety monitoring, providing timely feedback to the instrument operator; 9) data reduction tools and pipeline Level 1 -- 4 science product generation; 10) data archival following all SOFIA Data Cycle System (DCS) requirements; 11) user documentation. The HIRMES software architecture consists of two sub-systems: the HIRMES Command and Data Handling (CDH) software system, and the Data Analysis and Products (DAP) software. 

The HIRMES CDH software is based on the Aurora application framework (Steve~Maher, {\it private communication}), which provides flexible, platform-independent, Java-based utilities for scientific data systems and devices. Science data is archived by converting the detector data, telescope settings and astrometry, and instrument configuration into Flexible Image Transport System (FITS) format files. The detector data and telescope astrometry are tightly synchronized using the SOFIA Inter-Range Instrument Group (IRIG) timing framework. Indexes generated from the FITS header information permit fast retrieval and reporting of science/engineering data. The time-series FITS output of Aurora constitutes Level-1 data, and is archived in its entirety through the DCS. 

The Level-1 FITS data is then fed into a modified version of Comprehensive Reduction Utility for SHARC-2 \citep[CRUSH\footnote{CRUSH: \href{https://github.com/attipaci/crush}{https://github.com/attipaci/crush}};][]{2008SPIE.7020E..1SK}, which produces Level-2 data after performing the following steps: evenly spectrally \& spatially resampled (FITS data cube: RA, DEC, wavelength); re-orientated to North-up, East-left; instrument calibrated (removal of correlated noise, pixel masking, biasing, etc); and WCS \& DCS compliance. The DAP pipeline can then take this Level-2 data and produce Level-3 data after flux calibration and telluric correction. After this has been achieved, DAP can produce Level-4 data by combining data from multiple observations, stitching map-pointings together, extracting 2D or 1D spectra, etc. 

\subsection{Data Formats \& Access} 

All of the Level-1 to Level-4 data will be provided in multi-extension FITS data cubes of flux, variance, and instrument \& observation parameters. All data are fully FITS \& DCS compliant. Additionally, the data will be compliant with the NASA/IPAC InfraRed Science Archive (IRSA\footnote{IRSA:  \href{https://irsa.ipac.caltech.edu/}{https://irsa.ipac.caltech.edu/}}), to support plans for the DCS archive to be injested therein, within the next year or so. Viewing and exploration of these FITS data cubes will be possible via a future modified version of SpexTool\footnote{SpexTool: \href{http://irtfweb.ifa.hawaii.edu/~spex/observer/}{http://irtfweb.ifa.hawaii.edu/$\sim$spex/observer/}}, in addition to your local, friendly data cube viewers, such as QFitsView\footnote{QFitsView: \href{http://www.mpe.mpg.de/~ott/QFitsView/}{http://www.mpe.mpg.de/$\sim$ott/QFitsView/}}.  


\section*{Acknowledgments}

The development of HIRMES is funded by the NASA / SOFIA 3$^{rd}$ Generation Instrument solicitation to the NASA Goddard Space Flight Center and partnering institutions. We thank the USRA/SOFIA Science Operation Center staff and NASA Armstrong B703 staff for their on-going support during the development of HIRMES and its up-coming commissioning. 

This research was conducted [in part] at the SOFIA Science Center, which is operated by the Universities Space Research Association under contract NNA17BF53C with the National Aeronautics and Space Administration.

We would like to extends our thanks to David Franz, Kevin Denis, Manuel Balvin, George Manos, and Elissa Williams for their contribution towards useful discussions and fabrication support on the detector arrays.


\end{document}